\documentclass[jcp,aps,showpacs,supergroupedaddress,epsfig,amsmath,amssymb,floatfix,fleqn]{revtex4}
\usepackage{epsfig,amsmath,amsthm,amssymb,bm,epsf,graphics,psfrag,verbatim,subfigure,framed}
\usepackage[makeroom]{cancel}
\usepackage{hyperref}
\usepackage{wasysym}
\usepackage{mathtools}
\usepackage[usenames,dvipsnames]{color}
\usepackage{empheq}
\usepackage{bbm}
\usepackage{mathrsfs}
\usepackage{amsfonts}
\usepackage{color} 
\usepackage{pbox}
\def\tsigma{\widetilde{\sigma}}

\def\bv{{\mathbf b}}

\newcommand{\be}{\begin{equation}}
\newcommand{\ee}{\end{equation}}
\newcommand{\ba}{\begin{eqnarray}}
\newcommand{\ea}{\end{eqnarray}}
\newcommand{\bw}{\begin{widetext}}
\newcommand{\ew}{\end{widetext}}
\newcommand{\Tr}{{\rm{Tr}\,}}

\newcommand{\Jv}{{\mathbf{J}}}
\newcommand{\jv}{{\mathbf{j}}}

\newcommand{\vv}{{\bf v}}

\newcommand{\Ev}{{\bf E}}

\newcommand{\Bv}{{\bf B}}
\newcommand{\Hv}{{\bf H}}

\newcommand{\Av}{{\bf{A}}}

\newcommand{\rv}{{\mathbf{r}}}

\newcommand{\kv}{{\bf k}}

\newcommand{\dv}{{\bf d}}
\newcommand{\Dv}{{\bf D}}

\newcommand{\Rv}{{\bm{R}}}

\newcommand{\bing}[1]{\textcolor{black}{#1}}

\begin{document}

\title{The Casimir effect in topological matter}
\author{Bing-Sui Lu$^1$}
\email{binghermes@gmail.com, bslu@ntu.edu.sg}
\affiliation{$^{1}$ Division of Physics and Applied Physics, School of Physical and Mathematical Sciences, Nanyang Technological University, 21 Nanyang Link, 637371 Singapore.
}
\date{\today}

\begin{abstract}
We give an overview of the work done during the past ten years on the Casimir interaction in electronic topological materials, our focus being solids which possess surface or bulk electronic band structures with nontrivial topologies, which can be evinced through optical properties that are characterizable in terms of nonzero topological invariants. The examples we review are three-dimensional magnetic topological insulators, two-dimensional Chern insulators, graphene monolayers exhibiting the relativistic quantum Hall effect, and time reversal symmetry-broken Weyl semimetals, which are fascinating systems in the context of Casimir physics, firstly for the reason that they possess electromagnetic properties characterizable by axial vectors (because of time reversal symmetry breaking), and depending on the mutual orientation of a pair of such axial vectors, two systems can experience a repulsive Casimir-Lifshitz force even though they may be dielectrically identical. 
Secondly, the repulsion thus generated is potentially robust against weak disorder, as such repulsion is associated with a Hall conductivity which is topologically protected in the zero-frequency limit. 
Finally, the far-field low-temperature behavior of the Casimir force of such systems can provide signatures of topological quantization.  
\end{abstract}

\maketitle

\section{Introduction}

Casimir/van der Waals forces~\cite{casimir1948} are forces induced by quantum and thermal fluctuations of the background radiation~\cite{lifshitz1955,boyer1974,lambrecht2006,genet2003,parsegian-book,bordag-book}, which typically give rise to non-negligible attraction between bodies separated by distances of the micron scale or smaller. The attraction leads to stiction (i.e., the permanent, nonreversible sticking of device components) in nano and micro electromechanical systems (NEMS/MEMS)~\cite{allen2005}, which has been a major motivation to find mechanisms by which the Casimir/van der Waals attraction can be reduced~\cite{lu2015,dean2016} or even made repulsive~\cite{boyer1974,munday2009,hoye2018}. 
We may see how an attractive Casimir force arises between a pair of coplanar dielectric plates separated by a vacuum gap. If we assume that each dielectric interface does not cause polarizations to mix (as would be induced by roughness or a Hall current), and that the medium is dielectrically isotropic, so that a wave can be resolved into independent s (or TE) and p (or TM) polarization modes,  
the Casimir-Lifshitz energy per unit area can be expressed via~\cite{lifshitz1955,parsegian-book,bordag-book} 
\be
E(d) = k_BT \sum\limits_{n=0}^{\infty} \! {}^\prime \int \frac{d^2\kv_\perp}{(2\pi)^2} \Tr \ln \left( \mathbb{I} - \Rv_1\cdot \Rv_2 \, e^{-2 q_n d} \right). 
\label{lif0}
\ee
Here $d$ is the separation between the opposing surfaces, $k_B$ is the Boltzmann constant, $T$ is the temperature, the prime on the sum denotes that the $n=0$ term should be multiplied by a factor of $1/2$, $\kv_\perp \equiv (k_x, k_y)$, $q_n \equiv (\xi_n^2 + k_\perp^2)^{1/2}$, $k_\perp^2 \equiv k_x^2+k_y^2$ is the square transverse wavevector, parallel to the surface, $\xi_n \equiv 2\pi n k_BT/\hbar$ is the so-called Matsubara frequency, and $\Rv_i$ is the reflectivity matrix of the $i$th plate, defined by 
\ba
\Rv_i
&=& 
\begin{pmatrix}
r_{ss,i} & 0 \\
0 & r_{pp,i} 
\end{pmatrix}. 
\ea 
The symbols $r_{ss}$ and $r_{pp}$  denote the reflection coefficients for s- and p-polarization modes, respectively. 
Although expressed for finite temperatures, the Casimir-Lifshitz energy can be obtained also for the case of zero temperature by making the replacement $2\pi(k_BT/\hbar){\sum_{n=0}^{\infty}}' \to \int_0^\infty d\xi$~\cite{bordag-book}. 
For a pair of identical dielectric plates, the log trace term in the energy (\ref{lif0}) decomposes simply to $\ln (1 - r_{ss}^2 e^{-2q_n d}) + \ln (1 - r_{pp}^2 e^{-2q_n d})$. Thus for every frequency mode, the logarithms each decay monotonically with decreasing separation $d$, implying that the force is attractive. 
More generally, a theorem by Kenneth and Klich~\cite{kenneth2006,bachas2007} ensures that if a pair of objects separated by a vacuum gap are related by reflection (or mirror) symmetry, the resulting Casimir force is always attractive. The argument relies on an assumption that the objects are made of reciprocal media, i.e., they do not break time reversal symmetry. 

Thus, to overcome the problem of stiction and generate repulsive Casimir forces, one would have to look for scenarios which evade some of the assumptions presupposed by the Kenneth-Klich theorem. 
One such scenario is to make the configuration dielectrically asymmetric, as occurs when the intervening medium between the two opposing media has a dielectric permittivity intermediate to its neighboring media~\cite{dzyaloshinskii1961}. This can give rise to repulsive Casimir forces and nanolevitation in real plane-parallel systems~\cite{esteso1,esteso2}.  
On the other hand, this scenario is of limited relevance for addressing the issue of stiction in NEMS/MEMS, as the intervening medium is typically the vacuum. 
A second scenario is also to consider another asymmetric configuration, involving a purely dielectric object and a purely magnetically permeable object separated by the vacuum gap. This can also result in Casimir repulsion~\cite{boyer1974,hoye2018}. 
A third scenario is to choose an intervening gap which breaks spatial inversion symmetry, this can happen if the gap is filled by a Faraday material. This scenario, proposed by Jiang and Wilczek~\cite{jiang2019}, leads to left and right-circularly polarized electromagnetic modes of the same frequency propagating inside the gap with different phase velocities. The resulting Casimir force can oscillate between attraction and repulsion with interseparation distance. 

Yet a fourth scenario, and one which is germane to our review topic, occurs by retaining the vacuum gap but using dielectric plates which break time reversal symmetry, for example plates with surface magnetization. 
The time reversal symmetry-breaking (TRSB) field ${{\bm\Omega}}$ (in this case the surface magnetization) gives rise to a Hall conductivity $\sigma_{xy}$ in the plane parallel to the opposing surfaces, for such a situation it is known that the Onsager-Casimir relations apply: 
$\sigma_{ij}({{\bm\Omega}}) = \sigma_{ji}(-{{\bm\Omega}})$~\cite{zubarev1974,fain1969}. 
If the response function is an odd function of ${{\bm\Omega}}$, 
we have $\sigma_{ij}({{\bm\Omega}}) = - \sigma_{ji}({{\bm\Omega}})$. The antisymmetric part of the conductivity tensor is thus related to TRSB, and can be represented by an axial vector~\cite{footnote:axialvector}; physically this represents the circulation of the Hall current, and the sign of the axial vector tells us about the orientation of the circulation. 
As we subsequently see, whether or not the Casimir force between two dielectrically identical TRSB surfaces can become repulsive depends on the relative orientation of their Hall axial vectors. 

The presence of the Hall conductivity causes the polarization modes to mix at the dielectric interfaces, consequently in addition to $r_{ss}$ and $r_{pp}$, there are also mixed polarization reflection coefficients $r_{ps}$ and $r_{sp}$. 
For this scenario, instead of Eq.~(\ref{lif0}) the Casimir-Lifshitz energy per unit area is now given by~\cite{lambrecht2006,lambrecht2008,marachevsky2017,genet2003,fialkovsky2018,marachevsky2019} 
\be
E(d) = k_BT \sum\limits_{n=0}^{\infty} \! {}^\prime \int \frac{d^2\kv_\perp}{(2\pi)^2} \Tr \ln \left( \mathbb{I} - \Rv_1' \cdot \Rv_2 \, e^{-2 q_n d} \right). 
\label{lif}
\ee
Here, $\Rv_1'$ and $\Rv_2$ are the reflectivity matrices of the left and right plates defined by 
\be
\Rv_1'
= 
\begin{pmatrix}
r_{ss,1}' & r_{sp,1}' \\
r_{ps,1}' & r_{pp,1}'
\end{pmatrix}, 
\,\,
\Rv_2
= 
\begin{pmatrix}
r_{ss,2} & r_{sp,2} \\
r_{ps,2} & r_{pp,2}
\end{pmatrix}, 
\ee 
where the symbol $r_{\alpha\beta}$ ($\alpha,\beta = s,p$) denotes the reflection coefficient for an incident wave in the $\alpha$ polarization mode to be reflected in the $\beta$ polarization mode.  
For the left plate one has to solve a boundary value problem in which the surface normal and the directions of the incident ray and reflected rays are reversed, and the prime on $\Rv_1'$ reminds us of that the reflectivity matrix thus obtained is not necessarily identical with $\Rv_2$ even though the plates may be identical. 
If $r_{ps} = r_{sp}$ (as is the case for the TRSB topological materials reviewed in this paper) the trace log term in Eq.~(\ref{lif}) takes the form 
$\ln\big(
1 - ((r_{pp})^2 + 2 r_{ps,1}' r_{ps,2} + r_{ss}^2) e^{-2q_n d} + ((r_{ps,1}')^2 - r_{pp} r_{ss}) ((r_{ps,2})^2 - 
    r_{pp} r_{ss}) e^{-4q_n d}
\big)$, i.e., the logarithm is no longer guaranteed to decay monotonically with decreasing distance (as was the case for a TRS system without polarization mixing). 
As we shall subsequently see, monotonicity is not broken if $r_{ps,1}'$ and $r_{ps,2}$ have the same magnitude and the same sign, but can be broken if they have opposite signs, thus giving rise to the possibility of Casimir repulsion between identical plates. 
We will also see that a change of sign can be effected via a change of direction of the TRSB field, which affords one a way to control the sign of the Casimir force.  

In this review, we look into the Casimir effect in axionic topological insulators, Chern insulators, graphene monolayers subjected to perpendicular magnetic fields, and time reversal symmetry-broken Weyl semimetals. 
These are examples of the so-called electronic ``topological matter", an umbrella phrase which broadly covers electronic systems with properties characterizable by nonzero topological invariants (such as the Chern number). 
These invariants are related to the existence of a nontrivial topology in the electronic band structure, and show up for instance as topologically quantized values of the static Hall conductivity. Thus the static value of the Hall conductivity is quite robust and is not changed by the presence of weak (and nonmagnetic) disorder, and signatures of the topological quantization can appear in the behavior of the Casimir and Casimir-Polder interactions. Furthermore, these systems are potential candidates for generating Casimir repulsion through the aforementioned fourth, TRSB, scenario. In quantum Hall graphene, the TRSB field is the applied magnetic field; for Chern insulators and magnetic topological insulators, it is the ferromagnetic polarization of the surface. 

To the best of the author's knowledge, there have been three reviews on related topics in the past, viz., Refs.~\cite{woods2016}, \cite{khusnutdinov2019} and \cite{woods2021}. Reference~\cite{woods2016} provides an overview of realizations of the Casimir effect in various materials, e.g., lipid membranes, metamaterials, photonic crystals and plasmonic nanostructures, with a subsection concisely summarising the state of research (as of 2016) on Casimir interaction in topological matter. Reference~\cite{khusnutdinov2019} reviews the Casimir effect in two-dimensional Dirac materials, mainly emphasizing graphene systems. Reference~\cite{woods2021} is perhaps closest in spirit to the present review, offering a succinct overview of the Casimir effect in electronic topological materials. It is thus timely to provide a more detailed and pedagogical review of the Casimir effect in topological matter, which is consistent in terms of both approach and notation (e.g., we use Gaussian units throughout).

\section{Axionic topological insulators}

In the late noughties, a class of three-dimensional insulators, known as topological insulators~\cite{fu2007,moore2007,roy2009,hasan2010,fu2006,franz2013,bernevig2013,shen2017}, was predicted to possess topologically protected conducting surface states even though the bulk of the solid is insulating, and these surface states cannot be removed by non-magnetic disorder perturbations. Examples of such insulators include ${{\rm Bi_2 Se_3}}$, ${{\rm Bi_2 Te_3}}$ and ${{\rm Sb_2 Te_3}}$~\cite{hasan2010}. The topological insulators are distinguished from the ``trivial insulators" by virtue of a topological invariant called the $\mathbb{Z}_2$ invariant (denoted by the symbol $\nu$), whose value depends on the topological character of the electronic band structure of the insulator, and does not change if the insulator is deformed as long as the deformations do not modify the topology of the band structure~\cite{footnote:Z2}. For a topological insulator, the invariant has the value of 1, whereas for a trivial insulator, it has the value of 0. Thus, if the topological insulator is next to a trivial insulator, one cannot smoothly deform the topological insulator into the trivial insulator by continuously varying the value of the topological invariant. For the topological invariant to change, the band-gap has close and re-open on going across across the interface, thus the interface is a conductor, characterized by gapless surface states. 
Such surface states  also occur at the interface between a topological insulator and the vacuum, as the vacuum can be regarded as a trivial insulator with a band-gap equal to the rest mass of an electron-position pair. 

The surface states can be gapped by ferromagnetizing the surface of the three-dimensional topological insulator, with the magnetization axis perpendicular to each surface. In this way, one obtains a three-dimensional magnetic topological insulator. Such ferromagnetization can be induced, for example, by doping a topological insulator (such as ${{\rm Bi_2 Te_3}}$) with transition metals (such as Cr or Vn). The ferromagnetization can arise via the RKKY interaction for Mn-doped ${{\rm Bi_2 Te_3}}$ or the van Vleck mechanism for Cr-doped ${{\rm Bi, Sb}_2 {\rm Te}_3}$~\cite{tokura2019}. 
With the ferromagnetization, a Hall current emerges which circulates around the magnetization direction, breaking the time reversal symmetry. 
For an odd integer number of surface states, the Hall conductivity of each surface is correspondingly given by the odd integer multiplied by a factor of $e^2/(2h)$. 
Magnetic topological insulators are predicted to exhibit certain exotic electromagnetic phenomena, e.g., an electric charge will see an image magnetic monopole, whilst the Kerr and Faraday rotations become topologically quantized~\cite{qi2013,qi2009}, and indeed some of these phenomena have already been experimentally confirmed~\cite{wu2016,okada2016,dziom2017}. 

\subsection{equations of axion electrodynamics}

To study the electromagnetic properties of three-dimensional (3D) magnetic topological insulators, a widely used approach is that proposed by Refs.~\cite{qi2008,qi2011}, which is based on axion field theory~\cite{wilczek1987,nenno2020}. 
In this approach, there is an extra term to the usual Maxwell electromagnetic action, viz.,
\be
S_A = \frac{e^2}{32\pi^2 \hbar c} \int \! d^3\rv \, dt \, \theta \, \epsilon^{\mu\nu\alpha\beta} F_{\mu\nu} F_{\alpha\beta}, 
\label{S-axion}
\ee
where $\epsilon^{\mu\nu\alpha\beta}$ is the completely antisymmetric tensor, $F_{\mu\nu}$ is the Maxwell field strength tensor, and $\theta$ is the so-called axion coupling strength. In its original context, the axion is a hypothetical pseudoscalar particle first proposed in quantum chromodynamics (QCD) to address the strong CP problem~\cite{peccei1977a,peccei1977b}, which has also been regarded as a candidate for cosmological dark matter (for a review, see Ref.~\cite{luzio2020}). In the context of topological insulators, the word ``axion" is adopted because of the mathematical similarity of the term~(\ref{S-axion}) to that describing cosmological/QCD axions; physically, however, the axion coupling in topological insulators is related to the existence of a surface quantum Hall effect.   
The action gives rise to the modified Maxwell equations: 
\begin{subequations}
\label{axion-eqns}
\ba
{\bm \nabla} \!\cdot\! \Dv &\!=\!& 4\pi \rho 
+ \frac{\alpha}{\pi} {\bm \nabla} \!\cdot\! ( \theta \Bv ), 
\,\,
{\bm \nabla}\!\cdot\!\Bv = 0,
\\
{\bm \nabla} \!\times\! \Hv &\!=\!& \frac{1}{c} \partial_t \Dv + \frac{4\pi}{c} \Jv 
- \frac{\alpha}{\pi} 
\big( {\bm \nabla} \theta \times \Ev + \frac{1}{c} ( \partial_t \theta ) \Bv \big), 
\label{m3qi}
\\
{\bm \nabla} \!\times\! \Ev &\!=\!& - \frac{1}{c} \partial_t \Bv. 
\label{m4qi}
\ea
\end{subequations} 
Here, $\alpha \equiv e^2 / (\hbar c) \approx 1/137$ is the fine structure constant, and $\rho$ and $\Jv$ are the external charge density and external current. 
Topological insulators whose optical properties are described by the modified Maxwell equations are also known as an axionic topological insulators. 

The equations can be re-expressed in a form resembling that of the conventional Maxwell equations, if we make use of the identities ${\bm \nabla} \theta \times \Ev = {\bm \nabla} \times (\theta \, \Ev) - \theta {\bm \nabla} \times \Ev$ and $(\partial_t \theta) \Bv = \partial_t (\theta\,\Bv) - \theta \partial_t \Bv$, apply Eq.~(\ref{m4qi}), and define the modified constitutive equations, $\Hv' = \Hv + (\alpha/\pi) \theta \Ev$ and $\Dv' = \Dv - (\alpha/\pi) \theta \Bv$. Equations~(\ref{axion-eqns}) become 
\begin{subequations}
\ba
{\bm \nabla} \!\cdot\! \Dv' &\!=\!& 4\pi \rho,  
\,\,
{\bm \nabla}\!\cdot\!\Bv = 0,
\\
{\bm \nabla} \!\times\! \Hv' &\!=\!& \frac{1}{c} \partial_t \Dv' + \frac{4\pi}{c} \Jv, 
\\
{\bm \nabla} \!\times\! \Ev &\!=\!& - \frac{1}{c} \partial_t \Bv. 
\ea
\end{subequations} 
The axion coupling strength $\theta$ can be used to distinguish a topological insulator from a trivial insulator: $\theta = (2n+1) \pi$ (where $n \in \mathbb{Z}$) inside a topological insulator, whereas $\theta = 0$ inside a trivial insulator. It turns out that $\theta$ is related to the previously mentioned $\mathbb{Z}_2$ invariant $\nu$: $e^{i\theta} = 1$ corresponds to $\nu = 0$ whilst $e^{i\theta} = -1$ corresponds to $\nu = 1$~\cite{qi2013}. 
If a semi-infinite 3D axionic topological insulator is to the left of a semi-infinite trivial insulator and separated from it by a planar interface, we can express $\theta_L(\rv) = (2n+1) \pi \Theta(- z)$ (where the subscript $L$ indicates that the topological insulator is to the left, $z$ is the direction perpendicular to the interface located at $z=0$, and $\Theta(z)$ is the Heaviside function). If we neglect the external current, Eq.~(\ref{m3qi}) becomes 
\be
{\bm \nabla} \!\times\! \Hv = 
\frac{1}{c} \partial_t \Dv
- \frac{\alpha}{\pi} ( \partial_z \theta_L(\rv) ) \hat{z} \times \Ev 
=
\frac{1}{c} \partial_t \Dv
+ (2n+1) \alpha \, \delta(z) \hat{z} \times \Ev 
\equiv 
\frac{1}{c} \partial_t \Dv 
- \frac{4\pi}{c} \sigma_{xy,L} \delta(z) \hat{z} \times \Ev, 
\ee
where $\hat{z}$ is a unit vector pointing in the positive z direction, and in the last equality, we have introduced the Hall conductivity $\sigma_{xy}$ of a two-dimensional surface. 
Thus we see that at the interface, there is a surface Hall conductivity half-integer quantized in units of $e^2/h$, viz., $\sigma_{xy,L} = - \theta_L (e^2/(2\pi h)) = - (2n+1)e^2/(2h)$, and $\partial_z \theta_L(\rv)$ is related to the surface Hall conductivity. The axion field theory thus captures the surface Hall conductivity of magnetic topological insulators and reproduces its half-integer quantization in the static limit. 
For a topological insulator to the right of the trivial insulator, $\theta_R(\rv) = (2n+1) \pi \Theta(z)$ (where $R$ indicates the topological insulator is to the right), correspondingly the Hall conductivity is expressed as $\sigma_{xy,R} = \theta_R (e^2/(2\pi h))$ (where the subscript $R$ indicates that the topological insulator is to the right). 
On the other hand, the sign of the Hall conductivity $\sigma_{xy}$ is related to the orientation of the Hall current (or the axial vector). Thus, for a pair of coplanar topological insulator surfaces facing each other, if $\theta_L = - \theta_R$ then the surface Hall conductivities are equal in magnitude and the Hall current axial vectors are parallel;
conversely, if $\theta_L = \theta_R$ then the axial vectors are antiparallel. 
As we shall see later, the former configuration can result in Casimir-Lifshitz repulsion, whereas the latter always results in attraction. 

The reflection coefficients for an interface between an isotropic axionic medium and the vacuum can be obtained by solving a boundary value problem. For the case where the  topological insulator (vacuum) is to the right (left) of the interface, and the incident ray propagates to the right, the reflection coefficients are found to be~\cite{grushin2011a,grushin2011b,chang2009} 
\ba
r_{ss} &\!\!=\!\!& 
\frac{1}{\Delta} \big( 1 - \varepsilon - \bar{\alpha}^2 + \sqrt{\varepsilon} \chi_- \big),
\nonumber\\
r_{ps} &\!\!=\!\!& r_{sp} = 
\frac{2\bar{\alpha}}{\Delta}, 
\nonumber\\
r_{pp} &\!\!=\!\!& 
\frac{1}{\Delta} \big(  - 1 + \varepsilon + \bar{\alpha}^2 + \sqrt{\varepsilon} \chi_-  \big),  
\label{r-coeffs-T}
\ea
where $\bar{\alpha} \equiv \alpha\theta/\pi$, $\theta$ is the axion parameter, being equal to $0$ for a trivial insulator and an odd integer multiplied by $\pi$ for a topological insulator; $\varepsilon$ is the dielectric permittivity of the topological insulator slab, and 
\ba
\Delta &=& 1 + \varepsilon + \bar{\alpha}^2 + \sqrt{\varepsilon} \chi_+, 
\\
\chi_\pm &=& \frac{\xi^2 + c^2 k_\perp^2 \pm \big(\xi^2 + \frac{c^2 k_\perp^2}{\varepsilon}\big)}{\sqrt{(\xi^2 + c^2 k_\perp^2) \big(\xi^2 + \frac{c^2 k_\perp^2}{\varepsilon}\big)}}.  
\ea
Here, $k_\perp = |\kv_\perp|$ where $\kv_\perp = (k_x, k_y)^{{\rm T}}$ is a transverse wavevector, and $\xi$ is the magnitude of the imaginary frequency. 
For the case where the topological insulator is to the left of the interface and the vacuum is to the right, and the incident ray propagates to the left, the reflection coefficients remain the same, i.e., $r_{\alpha\beta}' = r_{\alpha\beta}$~\cite{footnote:sign}.


\subsection{Casimir force behavior}

The behavior of the Casimir-Lifshitz force between a pair of infinitely thick axionic topological insulators with opposing planar surfaces separated by a vacuum gap is studied in Ref.~\cite{grushin2011a,grushin2011b}. For separation distances between magnetic topological insulators of the order of microns, the effect of parasitic magnetic forces is much smaller than the Casimir-Lifshitz force and can be disregarded~\cite{bruno2002}. 
The Casimir-Lifshitz interaction energy is obtained by substituting the reflection matrix formed from Eqs.~(\ref{r-coeffs-T}) into Eq.~(\ref{lif}). Here it is assumed that whilst the dielectric functions are dispersive, the axionic coupling strengths are not. 
The behavior at zero temperature was studied in Ref.~\cite{grushin2011a} for the case of ${{\rm TlBiSe_2}}$, where a Lorentz oscillator model is adopted for the dielectric permittivity with a single oscillator, with a single resonant frequency at around $56 \, {{\rm cm}}^{-1}$ and the static permittivity is $\varepsilon(0) \approx 4$. 

For $\bar{\alpha}_i < 1$ (where $i=1$ refers to the topological insulator slab to the left of the vacuum gap, and $i=2$ refers to the topological insulator slab to the right), it is found that the Casimir-Lifshitz force is always attractive if the axionic coupling strengths $\bar{\alpha}_1$ and $\bar{\alpha}_2$ have the same sign, but can become repulsive if they have opposite signs~\cite{footnote:largeaxion}. 
For the latter case, the force is attractive at large separations, but becomes repulsive at sufficiently short separations. To see how the repulsion arises, we adopt an argument used in Ref.~\cite{grushin2011a}. Let us rewrite the trace log term in Eq.~(\ref{lif}) as a log det term, i.e., 
\ba
&&\ln \det \left( 1 - \Rv_1' \cdot \Rv_2 \, e^{-2q_n d} \right) 
\nonumber\\
&=& 
\ln \big(
1 
- (r_{pp1} r_{pp2} + r_{ps2} r_{sp1}' + r_{ps1}' r_{sp2} + r_{ss1} r_{ss2}) e^{-2q_n d} 
+ (r_{ps1}' r_{sp1}' - r_{pp1}' r_{ss1}') (r_{ps2} r_{sp2} - r_{pp2} r_{ss2}) e^{-4q_n d}
\big)
\ea
At zero temperature, the Casimir-Lifshitz energy involves an integral over the imaginary frequency $i\xi$ instead of a Matsubara sum. As one can change the integration variables in the Casimir-Lifshitz energy from $\xi$ and $\kv_\perp$ to $\tilde{\xi} \equiv \xi d/c$ and $\tilde{\kv}_\perp \equiv \kv d$, the dielectric permittivity in the one-oscillator model $\varepsilon(i\xi) = 1+\omega_e^2/(\xi^2+\omega_R^2+\gamma_R \xi)$ (where $\omega_e^2$ accounts for the oscillator strength, $\omega_R$ the resonance frequency and $\gamma_R$ the damping parameter) can be rewritten as $\varepsilon(i\tilde{\xi} c / d) = 1+\omega_e^2/((\tilde{\xi} c/d)^2+\omega_R^2+\gamma_R (\tilde{\xi} c/d))$. This implies that  as $d$ becomes small, the second term in $\varepsilon(i\tilde{\xi} c/d)$ starts decaying as $d^{2}$, and in the limit $d\to 0$, $\varepsilon(i \tilde{\xi} c/d) \to 1$ (dielectric transparency) and $\chi_- \approx 0$. From Eqs.~(\ref{r-coeffs-T}) we see that the reflection coefficients become 
$r_{ss,i} \approx - \bar{\alpha_i}^2/(2+\bar{\alpha_i}^2 + \chi_+)$, 
$r_{pp,i} \approx \bar{\alpha_i}^2/(2+\bar{\alpha_i}^2 + \chi_+)$, 
$r_{ps,1}' = r_{sp,1}' \approx 2\bar{\alpha_1}/(2+\bar{\alpha_1}^2 + \chi_+)$, 
and $r_{ps,2} = r_{sp,2} \approx 2\bar{\alpha_2}/(2+\bar{\alpha_2}^2 + \chi_+)$. 
The sign of the mixed polarization reflection coefficients is thus dictated by the sign of the axion coupling strength. 
Furthermore, $r_{ps}$ and $r_{sp}$ are dominant over $r_{ss}$ and $r_{pp}$ by an order of $\alpha$ (the fine structure constant). We can thus neglect $r_{ss}$ and $r_{pp}$, and the log det term then becomes approximately 
\be
\ln \det \left( 1 - \Rv_1' \cdot \Rv_2 e^{-2q_n d} \right) 
\approx 
\ln \big(
1 
- 2 r_{ps1}' r_{ps2} e^{-2q_n d} 
+ (r_{ps1}')^2 r_{ps2}^2 e^{-4q_n d}
\big) 
= 
2\ln\big( 1 - r_{ps1}' r_{ps2} e^{-2q_n d} \big).
\ee
Thus, if $\bar{\alpha}_1$ and $\bar{\alpha}_2$ have opposite signs, which corresponds to the Hall current axial vectors pointing in the same direction, $r_{ps1}'$ and $r_{ps2}$ would also have opposite signs, and the Casimir-Lifshitz energy increases with a decrease of $d$ for sufficiently small separations, correspondingly the force is repulsive. 
The Casimir-Lifshitz force changes sign if the separation distance is larger than a certain threshold value $d_{eq}$. The value of $d_{eq}$ increases with the value of the axion coupling strength, but becomes smaller for larger slab dielectric permittivities~\cite{grushin2011a}. 

Crucially, the foregoing argument depends on the assumption that the axion coupling strength is nondispersive, i.e., it is nonzero even at infinitely large frequency. A more realistic model taking frequency dispersion into account (such as that described in Ref.~\cite{dejuan2012}) leads to $\bar{\alpha}(i\tilde{\xi}c/d)$ decaying as $(d/(\tilde{\xi}c))^{2}$ as $d\to 0$, thus both diagonal (dielectric permittivity) and off-diagonal (axion) contributions to the reflectivity matrix would need to be taken into account~\cite{marachevsky2019} for a proper determination of the Casimir-Lifshitz force. The vanishing of the off-diagonal contribution can cause the force to revert to attraction at very short separations. 

The effect of finite temperature on the Casimir-Lifshitz force between semi-infinite topological insulator slabs is subsequently studied in Ref.~\cite{grushin2011b}. In particular, the high-temperature limit is explored, where repulsion is found to be still possible in principle, albeit for a much reduced dielectric permittivity (with a value smaller than 2) and sufficiently large axionic coupling strength. For example, for $\theta = 5 \pi$, $\varepsilon$ would have to be smaller than 0.2 for the force to become repulsive. 

In the foregoing, we have described the Casimir force behavior for semi-infinite slabs. 
The effect of finite thickness has been studied in Ref.~\cite{nie2013}, where it is found that if the slabs have finite thickness and are bounded only by the vacuum, reducing the slab thickness also leads to a larger value for $d_{eq}$; conversely, if the left topological insulator slab is bounded to the left by an infinite silicon substrate medium, and the right topological insulator slab is similarly bounded by an infinite silicon substrate, then the value of $d_{eq}$ undergoes a decrease. These agree with the expectation that the effect of having more dielectric matter tends to increase the strength of Casimir attraction. 

The Casimir force between two multilayered topological insulator structures separated by a vacuum gap has also been studied~\cite{zeng2016}. For the case where the surface magnetization of each layer points in the same direction, increasing the number of layers tends to increase $d_{eq}$ and also increases the strength of the repulsive Casimir force, when compared to the case of a pair of semi-infinite topological insulator slabs with the same surface magnetizations.

\subsection{van der Waals torque between birefringent topological insulators}

\begin{figure}[h]
\centering
  \includegraphics[width=0.27\textwidth]{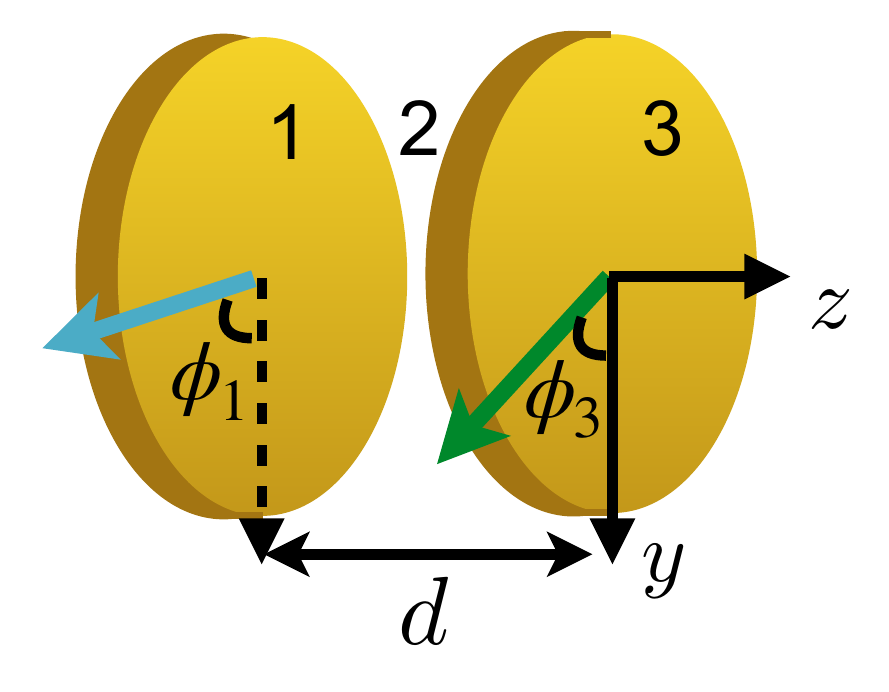}
  \caption{A pair of flat, coplanar topological insulator slabs (labeled 1 and 3) separated by a vacuum gap (labeled 2) of width $d$. The optic axis of slab 1 (3) is colored cyan (green), and the orientations of the optic axes measured with respect to a reference axis (in this case the $y$ axis) are $\phi_1$ and $\phi_3$. In the figure we have only shown a finite slice of the slabs, which are assumed to have infinitely large thicknesses and cross-sectional areas. Figure adapted from Ref.~\cite{lu2018}.} 
  \label{fig:slabs}
\end{figure}
Crystalline materials are not dielectrically isotropic in general, i.e., the dielectric permittivity of a crystal tends to have different values along different symmetry (or principal) axes of the medium. The effects of uniaxial dielectric anisotropy (i.e., where the dielectric permittivity of the bulk medium is the same along two principal axes but different along the third, the third axis being called the ``optic axis") have been studied in the retardation case for a configuration where the optic axes of the two slabs are both aligned with the normal to the topological insulator surface~\cite{grushin2011b}, and also in the non-retardation/van der Waals limit ($c \rightarrow \infty$) for a configuration where the optic axes are each perpendicular to the surface normal and also mutually misaligned (see Fig.~\ref{fig:slabs})~\cite{lu2018}. 
For the first configuration where the optic axes are aligned with the surface normals, it is found in Ref.~\cite{grushin2011b} that the Casimir repulsion can be enhanced by increasing the dielectric permittivity along the normal direction and/or reducing the dielectric permittivity along the transverse principal directions. 

For the configuration in which the optic axes are oriented perpendicular to the surface normal, two possibilities emerge: firstly, it becomes possible to tune the force between attraction and repulsion by changing the misalignment angle between the optic axes of the slabs, and secondly, a van der Waals torque also appears concurrently with the van der Waals force~\cite{parsegian1972,barash1979,lu2016,broer2019}. Such a torque arises owing to a dielectric mismatch in the azimuthal direction, in the same way that a force arises due to a dielectric mismatch in the surface normal direction. The behavior of the torque has been studied in the non-retardation regime~\cite{lu2018}, though not in the retardation regime. 
For the case of retardation, an extra complication arises due to the presence of an extraordinary wave contribution (in addition to the ordinary one) to the Casimir-Lifshitz energy~\cite{barash1979}. 

In what follows, we summarise the main results of Ref.~\cite{lu2018}. In the non-retardation regime, the Casimir-Lifshitz energy per unit area (effectively the van der Waals energy per unit area) is given by
\be
G(d) = k_{{\rm B}}T \, {\sum_{n=0}^{\infty}}' 
\!\int_0^{2\pi} \!\! \frac{d\psi}{2\pi} 
\!\int_0^\infty \!\! \frac{dQ}{2\pi} \, Q 
\ln \mathcal{D}(i\xi_n, Q d),   
\label{G}
\ee
where the function $\mathcal{D}$ is given by 
\ba
\mathcal{D}(\omega, Qd) 
&\!\equiv\!& 
1 - 
\frac{u_1 u_3 + 8 \, \bar{\alpha}_1 \, \bar{\alpha}_3 + \bar{\alpha}_1^2 \, \bar{\alpha}_3^2}{v_{1} v_{3}} 
e^{- 2 Qd} 
+
\frac{\bar{\alpha}_1^2 \, \bar{\alpha}_3^2}{v_1 v_3} 
e^{- 4 Qd}, 
\label{D1}
\ea
\begin{subequations}
and functions $u_i$ and $v_i$ are defined by
\ba
u_i &\equiv& \bar{\alpha}_i^2 + 2(\epsilon_{iz} g_i(\phi_i) - 1), 
\label{ui}
\\
v_i &\equiv& \bar{\alpha}_i^2 + 2(\epsilon_{iz} g_i(\phi_i) + 1) 
\label{vi}.
\ea
\end{subequations}
Here, $i=1, 3$ respectively label the left and the right slabs (cf. Fig.~\ref{fig:slabs}), $\phi_i$ denotes the orientation of the optic axis of slab $i$ relative to a reference axis, and $\epsilon_{ix}$, $\epsilon_{iy}$ and $\epsilon_{iz}$ denote the values of the dielectric permittivities along the three principal directions of the TI. The functions $g_i = 
\sqrt{1 + \gamma_i \cos^2(\phi_i - \psi)}$ and $\gamma_{i} \equiv {\epsilon_{ix}}/{\epsilon_{iz}} - 1$ give a measure of the degree of uniaxial anisotropy. 

It is found that increasing the anisotropy (that is, the relative difference between the dielectric permittivities along the optic and non-optic axes) has the effect of reducing repulsion and increasing attraction. 
The force oscillates with the angle of misalignment, being most repulsive/least attractive (least repulsive/most attractive) for angular differences that are integer (half-integer) multiples of $\pi$. For a certain range of dielectric permittivity values (of the order 1), the angle of misalignment of the optic axes can also be used to tune the force between attraction and repulsion. 
Compared to the torque between normal dielectrically anisotropic slabs, the presence of the axion coupling has the effect of reducing the torque between topological insulator slabs, with the reduction being greater if the axion coupling strengths of the slabs have opposite signs.

\subsection{Casimir-Polder interaction between an atom and a topological insulator}

\begin{figure}[h]
\centering
  \includegraphics[width=0.9\textwidth]{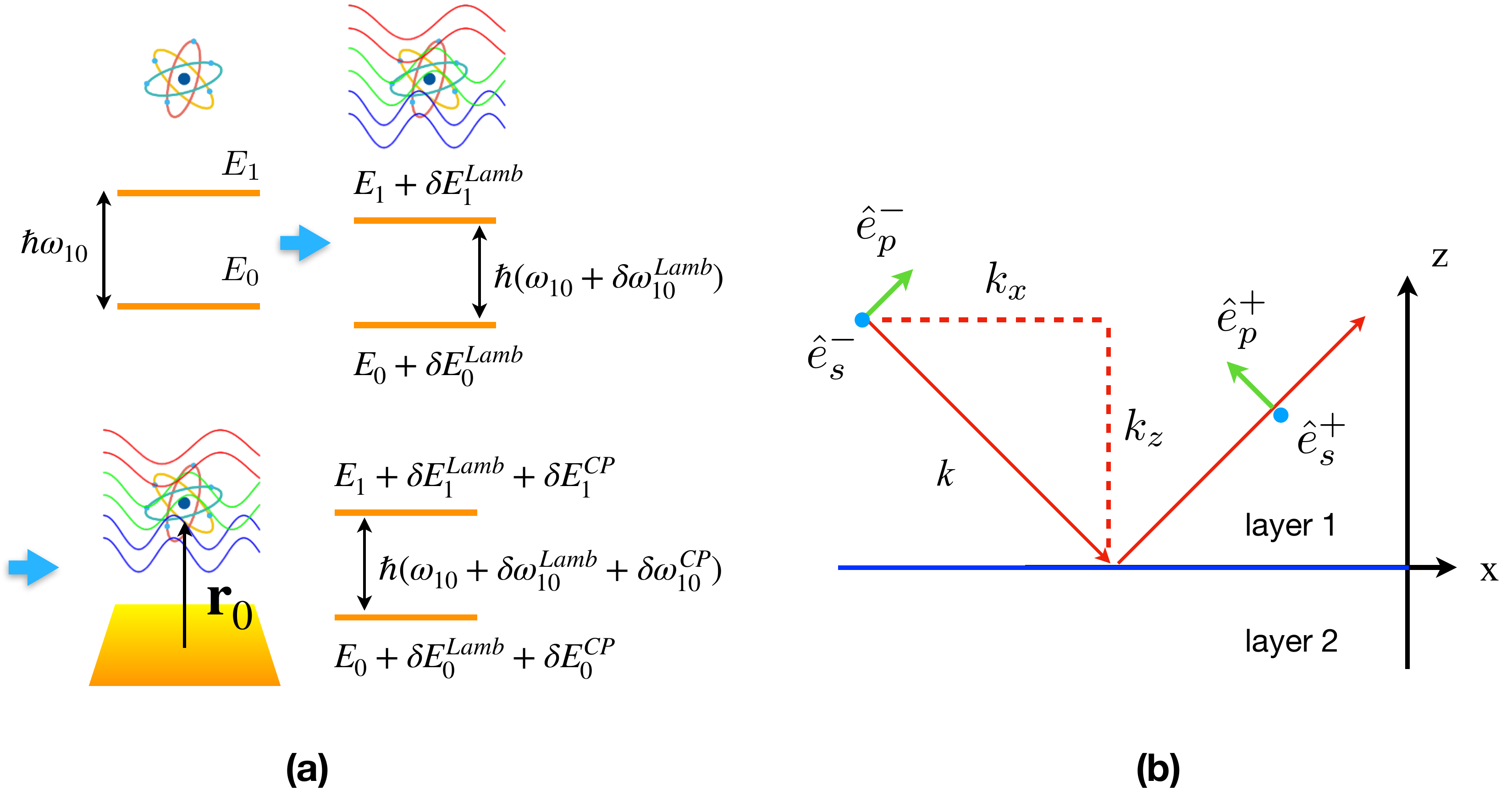}
  \caption{(a)~Schematic depiction of an atom (modeled by a two-level system with ground state energy $E_0$ and excited state energy $E_1$) in vacuum  interacting with an axionic topological insulator (represented by the gold surface). The interaction is via the dipole radiation coupling, in vacuum this gives rise to the Lamb shift $\delta E_a^{Lamb}$. As an atom in state $|a\rangle$ is brought near the surface, the surface induces an additional shift $\delta E_a^{CP}$, this is the Casimir-Polder energy for the atomic state $|a\rangle$. (b)~s and p polarisations, denoted respectively by $e_s^\pm$ (blue dot, directed along the negative y-direction) and $e_p^\pm$ (green arrow), for the case where the plane of incidence is the xz plane. For this case, the transverse wave vector lies entirely along the x direction, $k_y=0$, and $e_s^\pm = (0,-1,0)$ and $e_p^\pm = (1/k)(\mp k_z, 0, k_x)$. The $+$ ($-$) superscript refers to a wave propagating in the positive (negative) z-direction, $k = \omega/c$, and red arrows denote the propagation directions of incident and reflected waves.} 
  \label{fig:lamb}
\end{figure}
The study of the Casimir force between topological insulators leads naturally to the study of the Casimir-Polder (CP) interaction between a topological insulator and an atom. As is well-known, the CP interaction energy of a \emph{ground} state atom interacting with a dielectric slab can be obtained by the Lifshitz rarefaction method~\cite{lifshitz1955,bordag-book}, whereby one replaces the dielectric permittivity of one slab in the Casimir-Lifshitz formula~(\ref{lif}) with the sum of atomic polarizabilities of a dilute atomic gas, and then expand to leading order in the number density of atoms. The Casimir-Polder force $f^{CP}$ (acting in the $z$ direction, taken to be perpendicular to the surface which we assume to be planar) can be expressed by $f^{CP}(z) = - \partial \delta E^{CP}(z)/\partial z$, where $\delta E^{CP}$ is the Casimir-Polder energy.  
More generally, one can also study the CP energy of an \emph{excited} atomic state interacting with a dielectric slab (see Fig.~\ref{fig:lamb}a); to obtain this, one can calculate the shift induced in the energy levels of the atom by dipole radiation coupling in the presence of the dielectric slab via second-order quantum mechanical perturbation theory~\cite{wylie1984,wylie1985}. For an excited state, the CP energy ($E^{CP}$) has a so-called resonant contribution ($E^{CP,res}$), which is associated with the spontaneous emission of a real photon as the atom de-excites, and the photon is then absorbed by the dielectric medium. The use of the term ``resonant" is associated with the fact that the de-excitation frequency of the photon (corresponding to the energy difference between two atomic levels) matches an absorption frequency of the medium, so there is a resonant transfer of energy between the atom and the medium. The non-resonant contribution ($E^{CP,nres}$) is present in both excited and ground state atoms, and it comes from the exchange of virtual photons between the atom and the medium, which occurs at all frequencies. It has been shown experimentally for cesium atoms near a sapphire surface~\cite{failache1999,failache2003} that the near-field resonant CP force can be repulsive~\cite{gorza2001,gorza2006}, and furthermore this force is many times larger though much shorter-lived than the nonresonant CP force on a ground atomic state. 

Reference~\cite{fuchs2017} extends the investigation of both nonresonant and resonant CP energies to the case of an atom near a nondispersive axionic topological insulator at zero temperature. Specifically, they consider the case of an atom with two energy levels. 
If we generalize to a multilevel atom at position $\rv_0$ and label the possible atomic states by $|n\rangle$, the nonresonant Casimir-Polder energy of an atom in the state $|a\rangle$ can be put in the form~\cite{footnote-fuchs}
\ba
\label{ECP-nres}
&&\delta E_a^{CP,nres}(\rv_0) 
\\
&=&
- \frac{1}{\pi} 
\sum_{n \neq a}
\int_0^\infty \!\!\!\!\! d\xi \, 
d_\mu^{an} d_\nu^{na}
\left(
\frac{\omega_{na} \big( G_{\mu\nu}^R(\rv_0,\rv_0; i\xi) + G_{\beta\alpha}^{R\ast}(\rv_0,\rv_0; i\xi) \big)}{2(\omega_{na}^2 + \xi^2)}
+
\frac{\xi \big( G_{\mu\nu}^R(\rv_0,\rv_0; i\xi) - G_{\beta\alpha}^{R\ast}(\rv_0,\rv_0; i\xi) \big)}{2i(\omega_{na}^2 + \xi^2)}
\right). 
\nonumber  
\ea
$G_{\mu\nu}^R(\rv,\rv', \omega)$ is the reflection Green tensor for a source dipole  at $\rv'$ which is emitting a wave of frequency $\omega$, with the field point at $\rv$, $\mu, \nu = x,y,z$ label the directions in space, and $\dv_{an} \equiv \langle a | \hat{\dv} | n \rangle$ is a dipole transition matrix element, with $\hat{\dv} = (\hat{d}_x, \hat{d}_y, \hat{d}_z)$ being the electric dipole operator.
The first term describes the nonresonant CP energy for reciprocal materials, mathematically it comes from the coupling between the symmetric parts of the polarizability tensor and the reflection Green tensor. The second term only appears in systems which break Onsager reciprocity or equivalently, time reversal symmetry, and comes from the coupling between the antisymmetric parts of the polarizability tensor and the reflection Green tensor. 
The resonant CP energy of a multilevel atom at position $\rv_0$ is given by  
\be
\delta E_a^{CP,res} (\rv_0)
=
- \frac{1}{2} 
\sum_{n \neq a} 
d_\mu^{an} d_\nu^{na} 
\big( G_{\mu\nu}^R(\rv_0,\rv_0; \omega_{an}) 
+ G_{\beta\alpha}^{R\ast}(\rv_0,\rv_0; \omega_{an}) \big)
\Theta(\omega_{an}).
\label{ECP-res}
\ee
The reflection Green tensor itself is given by~\cite{gorza2001,crosse2015,lu2020}
\ba
{\mathbb G}^{R} (\rv, \rv_0; \omega)
&\!=\!& 
\frac{i}{2\pi} \left( \frac{\omega}{c} \right)^2 
\int d^2\kv_\perp 
\frac{1}{k_z}e^{i \kv_\perp \cdot (\rv_\perp - \rv_{0\perp}) + i k_{z} (z+z_0)} 
\nonumber\\
&&\times 
\big( r_{ss} \hat{e}_s^+ (\kv_\parallel)  \hat{e}_s^- (\kv_\parallel) 
+ r_{ps} \hat{e}_s^+ (\kv_\parallel) \hat{e}_p^- (\kv_\parallel)
+ r_{sp} \hat{e}_p^+ (\kv_\parallel) \hat{e}_s^- (\kv_\parallel)
+ r_{pp} \hat{e}_p^+ (\kv_\parallel) \hat{e}_p^- (\kv_\parallel) \big), 
\label{GR}
\ea
where $\kv_\perp = (k_x, k_y)^{{\rm T}}$, $k_z = ((\omega/c)^2 - k_\perp^2)^2$, 
$\rv_\perp = (x, y)^{{\rm T}}$, and 
$\hat{e}_\sigma^\pm$ ($\sigma = s, p$) are polarization vectors, shown in Fig.~\ref{fig:lamb}b. 

As found in Ref.~\cite{fuchs2017}, 
the oscillatory decay of the resonant CP energy shift for an atom of a given circular polarization near a topological insulator with an axion parameter value $\theta = \pi$ is antiphasal to the oscillatory decay of an atom of the same circular polarization near a topological insulator with $\theta = -\pi$. 
Similarly, one can deduce that the oscillatory decay in the case of a left-circularly polarized atom near a topological insulator of a given value of $\theta$ is antiphasal to that in the case of a right-circularly polarized atom near a topological insulator with the same value of $\theta$. 
Thus, the phase difference can help one to distinguish the circular polarization state of the excited atom or the sign of the axion parameter of the topological insulator. 
Changing the sign of the axion or the circular polarization direction of the atom also results in a change in the sign of the force. 

In Ref.~\cite{fang2019}, it is proposed that axionic topological insulators can be used to construct an atom trap. The idea is that of a cavity consisting of two surfaces, each of which is composed of a topological insulator and a negative refractive-index material (NRIM). The authors assume that the axion coupling strength can be made sufficiently large such that the reflection coefficients in Eq.~(\ref{r-coeffs-T}) become $r_{ss} \approx -1$, $r_{pp} \approx 1$ and $r_{ps} = r_{sp} \approx 0$, i.e., the topological insulator behaves effectively as a perfect conductor. For certain thicknesses of the NRIM, the cavity can exhibit a self-focusing ability, with a ray emitted from an atom placed inside the cavity undergoing negative refraction on reaching the vacuum-NRIM interface and a further internal reflection at the NRIM-TI interface. After a sequence of negative refractions and internal reflections, the ray eventually returns to the atom. This has an overall effect of suppressing the rate of spontaneous emission of the atom and prolonging the lifetime of the excited state. The resonant Casimir-Polder force would thus also be longer-lived. For such a cavity, the Casimir-Polder interaction has a potential well at the midpoint of the cavity, whose depth increases with the strength of the axion coupling and vanishes if the the axion coupling strength is zero. The cavity can thus be used for trapping cold atoms. 

In the above, we have described the Casimir-Polder interaction for an atom. 
The Casimir-Polder interaction is also being studied for molecules, in terms of experimental efforts at measuring the molecule-surface interaction~\cite{brand2015} as well as theoretical efforts at modelisation, such as that between a chiral molecule and a chiral metamaterial~\cite{butcher2012}. 
In the context of TRSB topological matter, the Casimir-Polder interaction between a molecule which violates the charge conjugation-parity symmetry (and thus also breaks time-reversal symmetry) and a TRSB material surface is investigated in Ref.~\cite{buhmann2018}. For this scenario, the Casimir-Polder interaction has an additional term which depends on the time-reversal odd-symmetric contributions to the polarizability tensor and the reflection Green tensor, which can be used to detect a putative violation of the charge conjugation-parity symmetry in molecules. 

\subsection{axion dispersion} 

The theoretical results described in the foregoing subsections have mainly assumed that the axion coupling strength is nondispersive. Such an assumption works well for the regime where inter-slab separations are sufficiently large such that the static limit becomes a good approximation. Thus the question arises as to how one should capture the frequency dispersion of the magnetic topological insulator for more general separations. 
To address this question, there have been proposed at least three approaches. One approach is via the use of five-dimensional QED~\cite{dejuan2012} (with the axion being the ``fifth" dimension), minimally coupling the gauge field to the (lattice) fermions, and calculating the polarization tensor by integrating out the fermions. The antisymmetric part of the polarization tensor yields a magneto-electric response which depends on the frequency, but not the wavevector. 

Another approach~\cite{chen2011,chen2012} is to model the gapped surface states of a magnetic topological insulator by the action of a massive Dirac fermion in $2+1$ dimensions and minimally coupling it to the gauge field. Again the polarization tensor can be obtained by integrating out the Dirac fermion, which causes the appearance of a Chern-Simons term in the effective electrodynamic action. They identify the coefficient in the Chern-Simons term as an effective axion response which depends on the wavevector and frequency, in addition to the surface band-gap, and find that there is a critical value of the band-gap above (below) which the system behaves like a topological (normal) insulator. In particular, the coefficient becomes the topologically quantized axion coupling strength as the band-gap becomes infinitely large. 

Finally, one can also regard a topological insulator as a system with surfaces characterized by frequency-dependent conductivity tensors~\cite{tse2010a,tse2010b,tse2011}. One starts with a Dirac Hamiltonian for the surface state and, having solved for the eigenstates, applies the Kubo formula to obtain the conductivity tensor. The antisymmetric part gives the Hall conductivity, which one interprets as being equivalent to the axion in terms of the axion coupling.

\section{Chern insulators}
\label{sec:CI}

By reducing the thickness of a magnetic topological insulator (such as Cr-doped ${{\rm Bi_2 Te_3}}$~\cite{chang2013}) until it becomes a film a few quintuple layers (or nms) thick, one obtains a Chern insulator~\cite{cayssol2013,weng2015,liu2016,ren2016,zhang2016}. A Chern insulator is effectively a two-dimensional insulator which exhibits the so-called quantum anomalous Hall effect at sufficiently low temperatures. This effect is characterized by the quantization of the static Hall conductivity to an integer multiple of $e^2/h$ (where $e = 1.6\times 10^{-19}$ C and $h$ is the Planck constant). The integer multiple is known as the Chern number $C$, a topological invariant expressing the winding number of a map from a two-dimensional torus to a two-dimensional unit sphere. 

The electromagnetic behavior of Chern insulators can be described by Maxwell's equations augmented by a surface conductivity contribution. In the frequency domain, the equations read~\cite{rodriguez-lopez2014}
\ba
{\bm\nabla} \!\cdot \Ev &\!\!=\!\!& 0, \,\,
{\bm\nabla} \!\cdot \Bv = 0, 
\nonumber\\
{\bm\nabla} \!\times\! \Hv 
&\!\!=\!\!& 
\frac{4\pi}{c} \delta(z){\bm{\sigma}}\cdot\Ev
- i \frac{\omega}{c} \Ev, 
\nonumber\\
{\bm\nabla} \!\times\! \Ev &\!\!=\!\!& i \frac{\omega}{c} \Bv, 
\label{maxwellCI}
\ea 
where the the two-dimensional conductivity tensor ${\bm\sigma}$ is given by 
\be
{\bm\sigma} =
\begin{pmatrix}
\sigma_{xx} & \sigma_{xy} & 0 
\\
\sigma_{yx} & \sigma_{yy} & 0
\\
0 & 0 & 0 
\end{pmatrix} =
\begin{pmatrix}
\sigma_{xx} & \sigma_{xy} & 0 
\\
- \sigma_{xy} & \sigma_{xx} & 0
\\
0 & 0 & 0 
\end{pmatrix}.
\ee
Here we have assumed the x and y directions to be in the plane of the Chern insulator. 
The second equality obtains on the assumption that the surface of the insulator is isotropic.

The corresponding reflection coefficients are obtained by solving a boundary value problem at the interface of the Chern insulator (cf. App.~\ref{app:1} for details). 
For the configuration shown in Fig.~\ref{fig:BCs2}(b), the reflection coefficients are  
\ba
r_{ss} &\!\!=\!\!& 
-\frac{1}{\Delta}
\big( 
\tsigma_{xx}^2 + \tsigma_{xy}^2
+ \frac{\omega}{c k_z} \tsigma_{xx} 
\big),
\,\,
r_{ps} = r_{sp} 
= \frac{1}{\Delta} \tsigma_{xy}, 
\nonumber\\
r_{pp} &\!\!=\!\!& 
\frac{1}{\Delta} 
\big( 
\tsigma_{xx}^2 + \tsigma_{xy}^2 
+ \frac{c k_z}{\omega} \tsigma_{xx}
\big), 
\ea
where $\tsigma_{\mu\nu}(\omega) \equiv (2\pi/c) \sigma_{\mu\nu}(\omega)$, $k_z \equiv ((\omega/c)^2 - k_\perp^2)^{1/2}$, and $\Delta \equiv \tsigma_{xx}^2 + \tsigma_{xy}^2 + \big( \frac{\omega}{c k_z} + \frac{c k_z}{\omega} \big) \tsigma_{xx} + 1$. 
Conversely, there is a change in sign for $r_{ps}'$ and $r_{sp}'$ (but not for $r_{pp}'$ and $r_{ss}'$) if the configuration in Fig.~\ref{fig:BCs2}(a) is considered instead. 
The above reflection coefficients indicate that we have to determine the frequency dispersion of the conductivity tensor if we are to properly characterize the Casimir interaction between Chern insulators. Thus, we shall turn to this problem in the next subsection.

\subsection{Surface conductivity via the Kubo formula}
 
A standard approach to calculate the conductivity tensor is to make use of a tight-binding model and apply the Kubo formula. 
Tight-binding lattice models have been constructed to realize different values of the Chern number $C$, e.g., $|C| = 1$, $2$ and $3$~\cite{qwz2006,grushin2012,sticlet2012,bernevig2013}. 
The simplest models are described by two-band Bloch Hamiltonians of the form
\be
\label{H2band}
H = {\bf{d}}(\kv_\perp) \cdot {\bm{\sigma}} = d_x(\kv_\perp) \sigma_x + d_y(\kv_\perp) \sigma_y + d_z(\kv_\perp) \sigma_z, 
\ee
where for a two-dimensional insulator, the wave-vector is two-dimensional and given by $\kv_\perp = (k_x, k_y)^{{\rm T}}$, and $\sigma_x, \sigma_y, \sigma_z$ are the Pauli matrices. 
The energy eigenvalues are 
\be
\varepsilon_{\pm}(\kv_\perp) = \pm d(\kv_\perp) = \pm \sqrt{d_x^2(\kv_\perp) + d_y^2(\kv_\perp) + d_z^2(\kv_\perp)}. 
\ee
Here, we have assumed that the chemical potential is zero, which is sufficient for the purpose of generating topological behavior, as long as the Fermi level is inside the gap. The chemical potential can always be controlled by tuning the gate voltage~\cite{chang2013}. 
In the Bloch sphere representation, the eigenstates are given by 
\ba
| +, \kv_\perp \rangle &\!\equiv\!& 
\begin{pmatrix}
\cos(\theta(\kv_\perp)/2) e^{-i\phi(\kv_\perp)}
\\
\sin(\theta(\kv_\perp)/2)
\end{pmatrix} 
e^{i\kv_\perp\cdot\rv}
\nonumber\\ 
&\!=\!& 
\frac{1}{2\sin(\theta(\kv_\perp)/2)}
\begin{pmatrix}
\sin \theta(\kv_\perp) e^{-i\phi(\kv_\perp)}
\\
1 - \cos \theta(\kv_\perp)
\end{pmatrix} 
e^{i\kv_\perp\cdot\rv},
\nonumber\\
| -, \kv_\perp \rangle &\!\equiv\!& 
\begin{pmatrix}
\sin(\theta(\kv_\perp)/2) e^{-i\phi(\kv_\perp)}
\\
- \cos(\theta(\kv_\perp)/2)
\end{pmatrix} 
e^{i\kv_\perp\cdot\rv}
\nonumber\\ 
&\!=\!& 
\frac{1}{2\cos(\theta(\kv_\perp)/2)}
\begin{pmatrix}
\sin \theta(\kv_\perp) e^{-i\phi(\kv_\perp)} 
\\
- 1 - \cos \theta(\kv_\perp)
\end{pmatrix} 
e^{i\kv_\perp\cdot\rv}, 
\label{eigenkets}
\ea
where $+$ ($-$) refers to the conduction (valence) band, $d_x/d = \cos\phi\sin\theta$, $d_y/d = \sin\phi\sin\theta$, $d_z/d = \cos\theta$, or equivalently, $\cos\phi = d_x/(d_x^2 + d_y^2)^{1/2}$, $\sin\phi = d_y/(d_x^2 + d_y^2)^{1/2}$, $\cos\theta = d_z/d$, and $\sin\theta = (d_x^2 + d_y^2)^{1/2}/d$. 
The eigenstates can also be written as 
\begin{subequations}
\ba
|+, \kv_\perp \rangle &\!\equiv\!& \sqrt{\frac{d(\kv_\perp)-d_z(\kv_\perp)}{2d(\kv_\perp)}} 
\begin{pmatrix}
\frac{d_x(\kv_\perp) - id_y(\kv_\perp)}{d(\kv_\perp)-d_z(\kv_\perp)}
\\
1
\end{pmatrix} 
e^{i\kv_\perp\cdot\rv}, 
\\
| -, \kv_\perp \rangle &\!\equiv\!& \sqrt{\frac{d(\kv_\perp)+d_z(\kv_\perp)}{2d(\kv_\perp)}} 
\begin{pmatrix}
\frac{d_x(\kv_\perp) - id_y(\kv_\perp)}{d(\kv_\perp)+d_z(\kv_\perp)}
\\
- 1
\end{pmatrix} 
e^{i\kv_\perp\cdot\rv}, 
\ea
\end{subequations}
where $\dv = (d_x, d_y, d_z)$. 
By working in a basis furnished by the above eigenstates, we can calculate the conductivity tensor by using the Kubo formula~(\cite{shen2017,czycholl2017}, cf. also App.~\ref{app:kubo}): 
\be
\sigma_{\mu\nu}(\omega) = 
-
\lim_{\delta\rightarrow 0} 
\frac{i\hbar}{\mathcal{A}}
\sum_{\ell,\ell'} 
\frac{\langle \ell | j_\mu | {\ell'} \rangle \langle {\ell'} | j_\nu | {\ell} \rangle}{\varepsilon_\ell - \varepsilon_{\ell'} + \hbar \omega + i\delta} 
\frac{f(\varepsilon_\ell) - f(\varepsilon_{\ell'})}{\varepsilon_\ell - \varepsilon_{\ell'}}, 
\label{kubo-general}
\ee
where $\mathcal{A}$ is the surface area, 
$f(\varepsilon_\ell) = (\exp(\beta \varepsilon_\ell) + 1)^{-1}$ is the Fermi-Dirac distribution, $j_\mu$ is the current operator, 
and $\sum_{\ell}$ denotes a sum over quantum states labeled by a set of quantum numbers collectively represented by the symbol $\ell$. 
For a two-band Chern insulator studied in a basis of Bloch states, 
$\ell = \{ \kv_\perp, n \}$ where $n = \pm$ is the band index. 
Thus, the above formula for the conductivity tensor becomes 
\be 
\sigma_{\mu\nu}(\omega) = 
-
\lim_{\delta\rightarrow 0} 
\frac{i\hbar}{\mathcal{A}}
\sum_{\kv_\perp \in BZ}
\sum_{n,n'=\pm} 
\frac{\langle n, \kv_\perp | j_\mu | n', \kv_\perp \rangle \langle n', \kv_\perp | j_\nu | n, \kv_\perp \rangle}{\varepsilon_n(\kv_\perp) - \varepsilon_{n'}(\kv_\perp) + \hbar \omega + i\delta} 
\frac{f(\varepsilon_n(\kv_\perp)) - f(\varepsilon_{n'}(\kv_\perp))}{\varepsilon_{n}(\kv_\perp) - \varepsilon_{n'}(\kv_\perp)}. 
\label{kubo-p}
\ee
\begin{figure*}[t]
\begin{center}
\begin{minipage}[b]{0.45\textwidth}
\includegraphics[width=\textwidth]{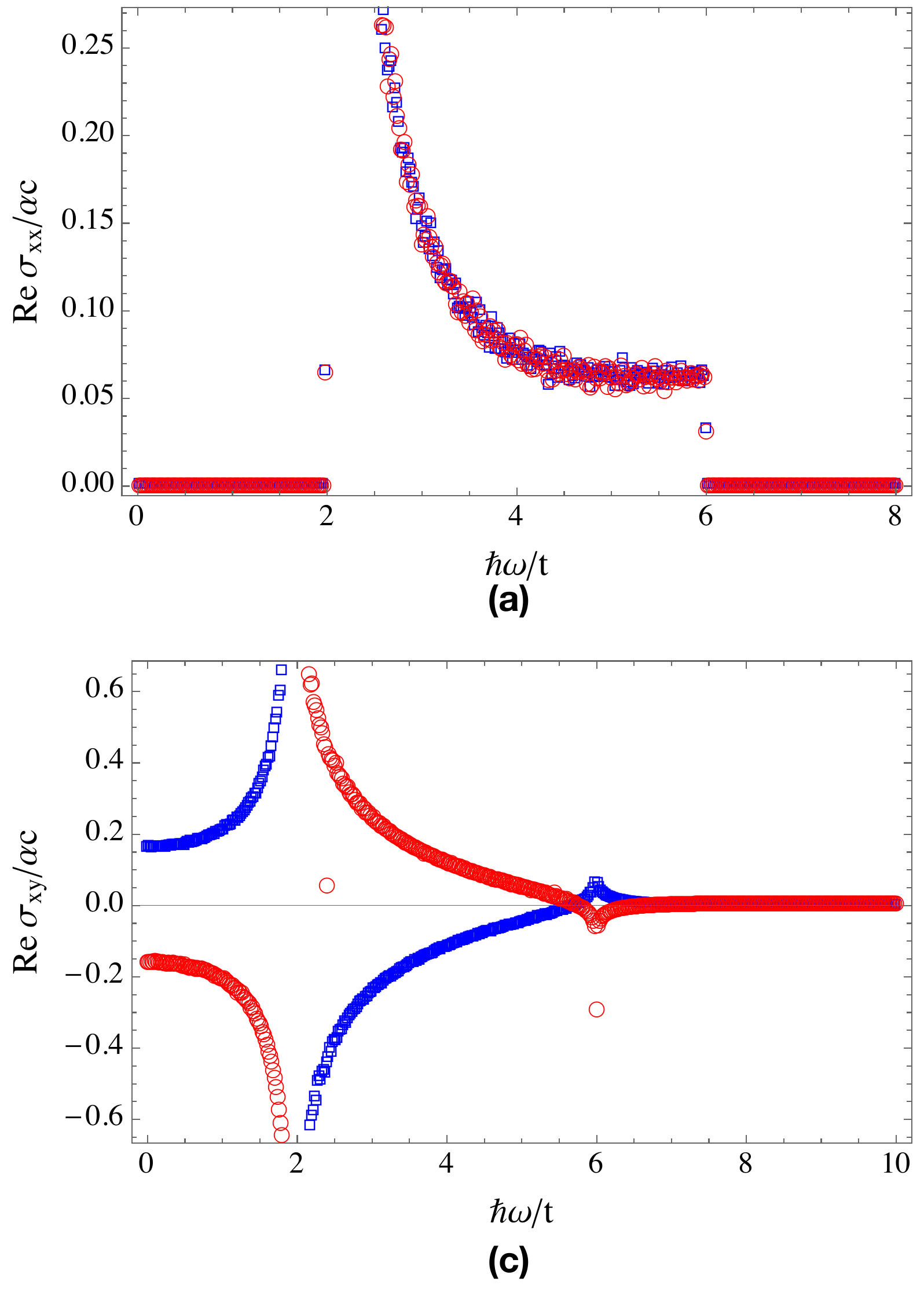} 
\end{minipage} 
\begin{minipage}[b]{0.45\textwidth}
\includegraphics[width=\textwidth]{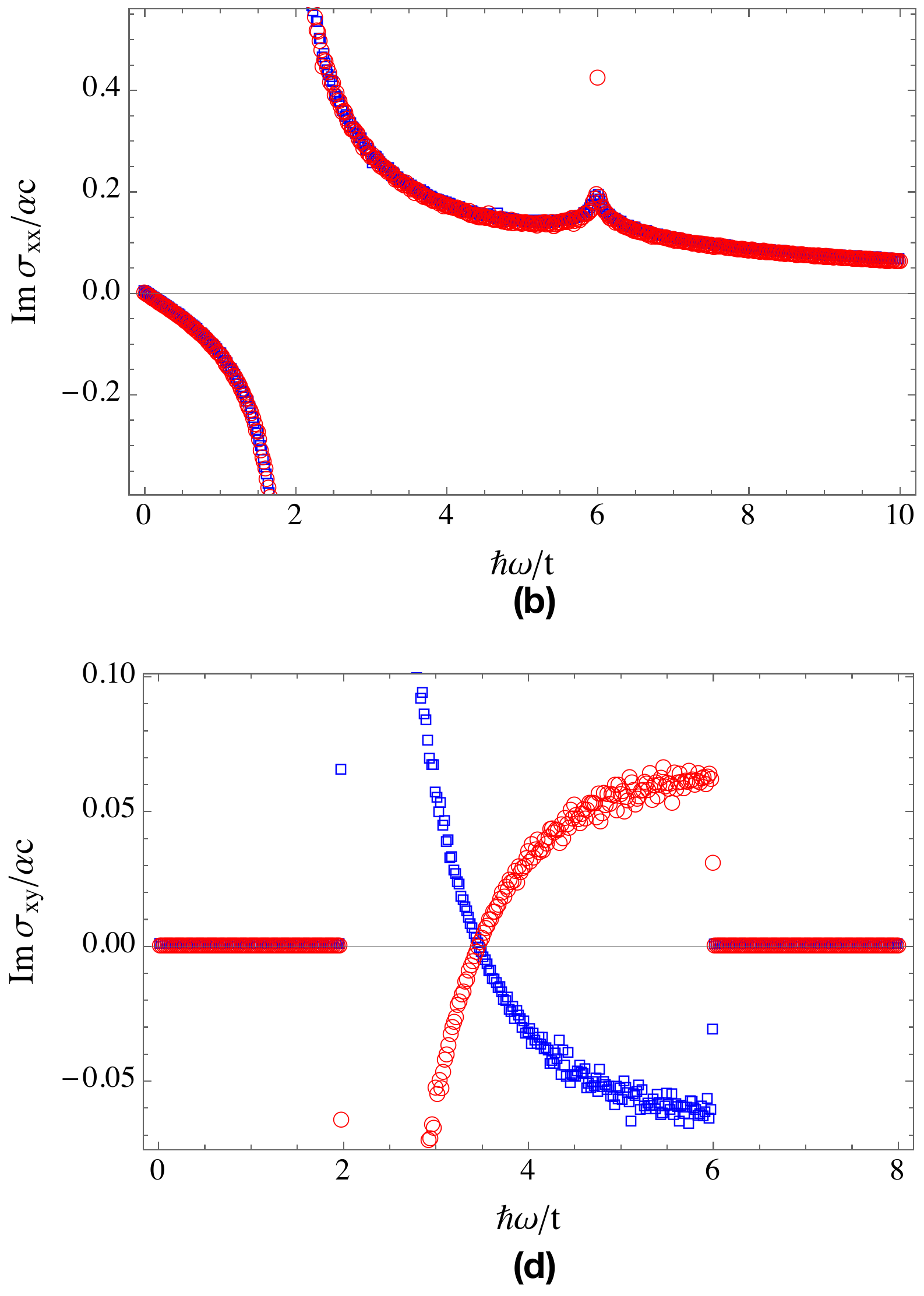} 
\end{minipage}
\end{center}
\caption{(color online) Real-frequency dispersion behavior, shown by curves with blue squares (red circles), of the Chern insulator's conductivity tensor for $u/t = 1$ ($u/t = -1$) in Eq.~(\ref{qwz}): (a)~${{\rm Re}} \, \sigma_{xx}(\omega)/(\alpha c)$, (b)~${{\rm Im}} \, \sigma_{xx}(\omega)/(\alpha c)$, (c)~${{\rm Re}} \, \sigma_{xy}(\omega)/(\alpha c)$, and (d)~${{\rm Im}} \, \sigma_{xy}(\omega)/(\alpha c)$ as functions of $\hbar \omega/t$ (horizontal axis). Figure reproduced from Ref.~\cite{lu2020}.}
 \label{conduct-realC1}
 \end{figure*}
As an example of a Chern insulator with $|C| = 1$, we can consider the Qi-Wu-Zhang (QWZ) model~\cite{qwz2006}, this is defined on a square lattice with a lattice constant $a$, and described by the Bloch Hamiltonian~(\ref{H2band}) with
\ba
d_x(\kv_\perp) &\!=\!& t \sin k_x a, 
\,\,
d_y(\kv_\perp) = t \sin k_y a,
\nonumber\\
d_z(\kv_\perp) &\!=\!& t (\cos k_x a + \cos k_y a) + u. 
\label{qwz}
\ea
The symbol $t$ denotes the hopping parameter, and $u$ denotes the band gap. 
For this model, it is known that the Chern insulator is in a phase with $C = 1$ for $0 < u < 2t$, $C = -1$ for $-2t < u < 0$, and $C = 0$ for other values of $u$~\cite{bernevig2013}. 

The full frequency dispersion of the conductivity tensor in the QWZ model is studied in Ref.~\cite{lu2020}, and the behavior is shown in Fig.~\ref{conduct-realC1} for the case where $|u| = t$ (which induces a $|C| = 1$ phase). In particular, Fig.~\ref{conduct-realC1}c shows that at zero frequency the real part of the Hall conductance is quantized at $e^2/h$. 
More generally, the conductivity tensor of the Chern insulator in the zero frequency limit is given by
\ba
&&\sigma_{xy}(\omega=0) = - \sigma_{yx}(\omega=0) = Ce^2/h, 
\nonumber\\
&&\sigma_{xx}(\omega=0) = \sigma_{yy}(\omega=0) = 0,
\label{sigma-chern-static}
\ea
and the corresponding reflection coefficients are given by 
\ba
&&r_{ss}(\omega=0) = - r_{pp}(\omega=0) = - \frac{(C\alpha)^2}{1+(C\alpha)^2}, 
\nonumber\\
&&r_{sp}(\omega=0) = r_{ps}(\omega=0) = \frac{C\alpha}{1+(C\alpha)^2}. 
\label{r-chern-static}
\ea
The reflection coefficients in the static limit coincide with those found for a Chern-Simons surface in Ref.~\cite{marachevsky2017}, since the Chern-Simons surface is equivalent to a Chern insulator in the static limit (as we see in Sec.~\ref{sec:CS}). There is a reason for the quantization of $\sigma_{xy}(\omega=0)$. In the static limit, the Hall conductivity becomes proportional to a topological invariant called the Chern number, which is an integer given by~\cite{kohmoto1985} 
\be
C 
= \frac{1}{4\pi} \int_{BZ} \!\!\!\! d^2\kv_\perp 
\left(
\frac{\partial \dv}{\partial k_x} \times \frac{\partial \dv}{\partial k_y}
\right)
\cdot 
\frac{\dv}{|\dv|^3}.  
\label{chern-def1}
\ee
In the above, $\dv = (d_x, d_y, d_z)^{{\rm T}}$. 
The above equation leads to a mathematical as well as a physical interpretation. In the first (mathematical) interpretation~\cite{footnote:dubrovin}, $C$ is the winding number of a map from the two-dimensional Brillouin zone parametrized by $k_x$ and $k_y$ to the unit sphere $S^2$ swept out by the unit vector $\dv/|\dv|$; this is just a higher-dimensional analogue of the winding number $w$ of a map from a one-dimensional period (parameterized by $t$) to a unit circle $S^1$ swept out by the unit vector $\rv/|\rv|$, where $\rv(t) = (x(t),y(t))^{{\rm T}}$ and $w = (1/2\pi)\int dt ((xdy-ydx)/r^2)$. In the second (physical) interpretation~\cite{shankar2018}, we note that $(\partial \dv/\partial k_x) \times (\partial \dv/\partial k_y) k_x k_y = d\Av$ describes an area element on the surface swept out by $\dv$ as $k_x$ and $k_y$ are varied across the entire Brillouin zone. For the QWZ model described by Eqs.~(\ref{qwz}), such a surface is a two-dimensional torus. Thus, we can rewrite Eq.~(\ref{chern-def1}) as 
\be
C = \oiint d\Av \cdot \Bv,  
\ee
where $\Bv \equiv (1/4\pi) \dv/|\dv|^3$ can be regarded as the field emitted by a magnetic monopole (i.e., the Dirac point where $|\dv| = 0$) sitting at the origin in $\dv$-space. Then $C$ would be the net magnetic flux passing through the surface. This is nonzero if the surface encloses the magnetic monopole or Dirac point, and zero if it doesn't. Whether the surface encloses the monopole or not depends on its relative distance from the origin, and  for the QWZ model this distance is controlled by the parameter $u$.

\subsection{lattice models for higher Chern numbers} 

We have seen how one can make use of a tight-binding model to determine the dispersion in the conductivity response of a Chern insulator, via the Kubo formula. 
To generate $C = \pm 1$ topological phases, one can make use of the QWZ model. One can go further and construct models that generate higher Chern numbers, keeping in mind that for Chern insulators, $d_x$ and $d_y$ ($d_z$) have odd (even) parity, i.e., $d_x$ and $d_y$ ($d_z$) have to be odd (even) under the simultaneous interchanges $k_x \to - k_x$ and $k_y \to - k_y$. 
A model which gives rise to $C = \pm 1, \pm 2$ was constructed by Grushin, Neupert, Chamon and Mudry in Ref.~\cite{grushin2012}, the Bloch Hamiltonian is specified by 
\ba
d_x(\kv_\perp) &\!=\!& t \sin k_x a, 
\,\,
d_y(\kv_\perp) = t \sin k_y a,
\nonumber\\
d_z(\kv_\perp) &\!=\!& h_1 \cos k_x a + h_2 \cos k_y a + u + M \cos k_x a \cos k_y a. 
\label{C2CI}
\ea
If we choose $h_1 = h_2 = t$ and $M = 0$, we recover the QWZ model which induces topological phases with $C = \pm 1$. If we choose $h_1 = h_2 = u = 0$, the model gives rise to a phase with $C = \pm 2$, with the sign of $C$ determined by the sign of $M$, and the band-gap at the $\Gamma$ point $(k_x, k_y) = (0, 0)$ is given by $2M$. 

\begin{figure}[h]
 \centering
    \includegraphics[width=0.31\textwidth]{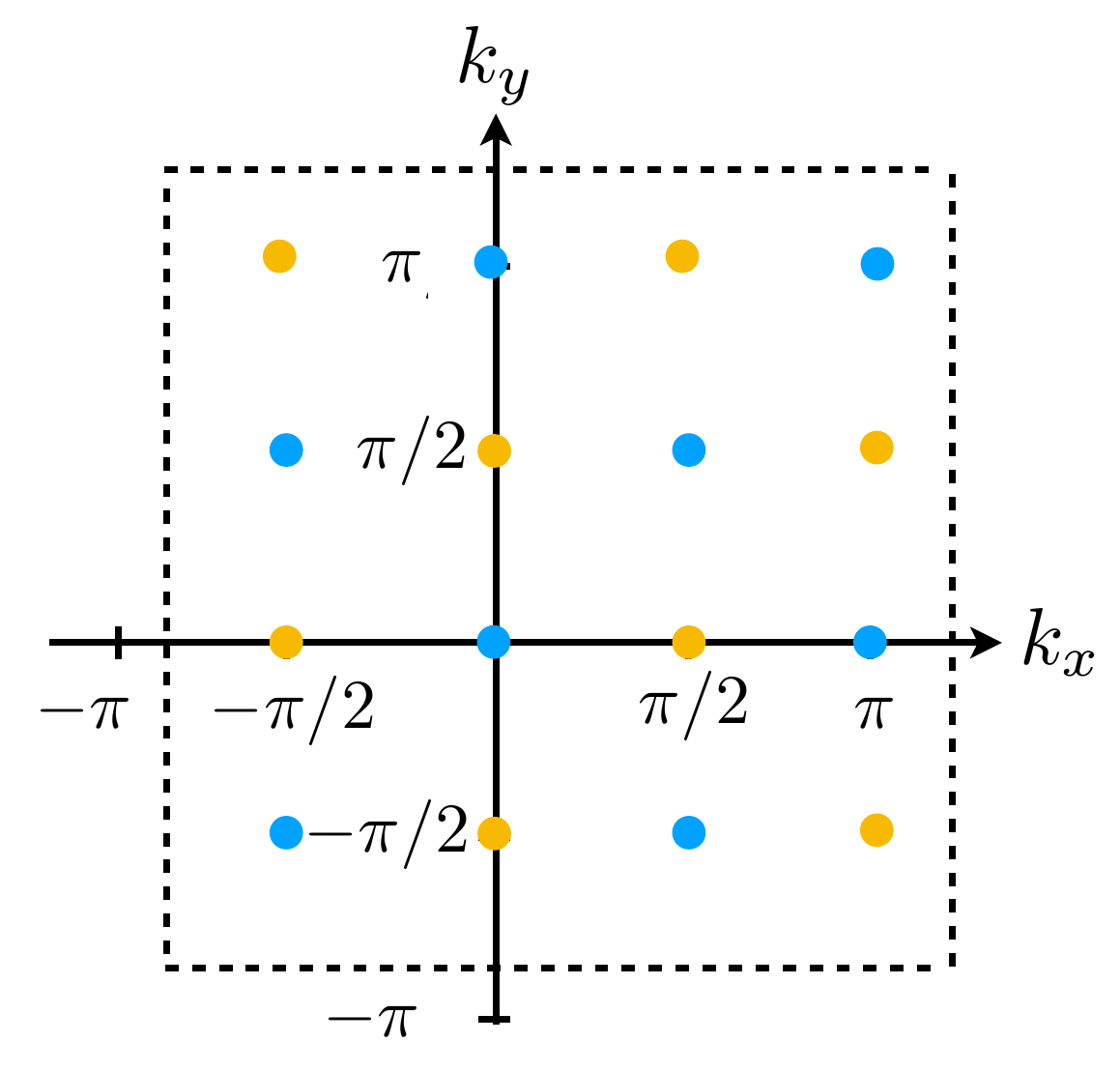}
     \caption{The first Brillouin zone for a Chern insulator described by Eq.~(\ref{C3CI}), showing the Dirac points enclosed. Dirac points with positive (negative) chirality are colored blue (ochre); cf. Table~\ref{table2}.} 
     \label{BZ2}
\end{figure} 
An example of a lattice model which can give rise to $C = \pm 1, \pm 3$ was proposed by Bernevig and Hughes in a homework problem in Ch.~8 of Ref.~\cite{bernevig2013}. Their Bloch Hamiltonian is specified by 
\ba
d_x(\kv) &\!=\!& t \sin 2 k_x a, 
\,\,
d_y(\kv) = t \sin 2 k_y a,
\nonumber\\
d_z(\kv) &\!=\!& M - t \cos k_x a - t \cos k_y a. 
\label{C3CI}
\ea
Here, the band-gap at the $\Gamma$ point $(k_x, k_y) = (0, 0)$ is $M - 2t$. 
The Dirac points are the positions $(k_x, k_y)$ in wavevector space where $d$ vanishes for appropriate values of $M$ (i.e., the valence and conduction bands touch and the band-gap closes). For example, for $M = 2t$, $d$ would be zero if $k_x = k_y = 0$, and thus $(k_x a, k_y a) = (0, 0)$ furnishes a Dirac point. For $M = t$, $d$ vanishes at $(k_x a, k_y a) = (\pi/2,0), (0, \pi/2), (-\pi/2,0), (0,-\pi/2)$, so $(\pi/2,0), (0, \pi/2), (-\pi/2,0), (0,-\pi/2)$ are also Dirac points. 
The Dirac points corresponding to various values of $M/t$ are listed in Table~\ref{table2}. 
Note that $M/t = 2, 1, 0, -1, -2$ are the only values of $M/t$ for which $d$ can vanish (which corresponds to the system becoming non-insulating). 
Topological insulator phases exist only when $d \neq 0$ (band-gap is open). Thus the phases of the Chern insulator are specified by $1< M/t < 2$, $0< M/t < 1$, $-1< M/t < 0$ and $-2< M/t < -1$. 

To calculate the Chern number, one may make use of the formula (\ref{chern-def1}), which is given in terms of a continuous integral over the Brillouin zone. 
On the other hand, there exists an equivalent formula which involves discrete summation rather than continuous integration; it is given by~\cite{sticlet2012} 
\be
C = \frac{1}{2} \sum_{\kv \in \Dv_i} {{\rm Sgn}}(\partial_{k_x}\dv\times\partial_{k_y} \dv)_z {{\rm Sgn}}(d_z). 
\label{chern-formula}
\ee
Here $\Dv_i$ denotes the $i$th Dirac point, and ${{\rm Sgn}}$ denotes the sign of the corresponding quantity. 
The quantity $(\partial_{k_x} \dv \times \partial_{k_y} \dv)_z$ is called the chirality, and we can define 
$J_i \equiv (\partial_{k_x} \dv \times \partial_{k_y} \dv)_z$ 
for the chirality at the $i$th Dirac point; this describes the orientation of the winding of $\dv$ at the given Dirac point.  
For the model described by Eq.~(\ref{C3CI}), $J_i = 4\cos 2k_x \cos 2k_y$. 
Using Eq.~(\ref{chern-formula}), we can calculate $C$ for different ranges of $M$; we list the values in Table~\ref{table3}. 
\begin{table}[hbtp]
\begin{tabular}{|c|l|c|}
\hline
$M/t$ & Dirac points $(k_x a, k_y a) \in \Dv_i$ & $J_i$ \\ \hline
$2$ & $(0,0)$ & $+$ \\ \hline
$1$ & $(\pi/2,0)$, $(0, \pi/2)$, $(-\pi/2,0)$, $(0,-\pi/2)$ & $-$ \\ \hline
$0$ & $(\pi/2,\pi/2)$, $(\pi/2,-\pi/2)$, $(-\pi/2,\pi/2)$, $(-\pi/2,-\pi/2)$, $(0, \pi)$, $(\pi, 0)$ & $+$ \\ \hline
$-1$ & $(\pi/2,\pi)$, $(\pi, \pi/2)$, $(-\pi/2,\pi)$, $(\pi,-\pi/2)$ & $-$ \\ \hline
$-2$ & $(\pi,\pi)$ & $+$ \\ \hline
\end{tabular}
\caption{Dirac points corresponding to different values of $M/t$, and the corresponding chiralities $J_i$.}
\label{table2}
\end{table}

\begin{table}[hbtp]
\begin{tabular}{|l|c|c|c|c|}
\hline
Dirac points $(k_x a, k_y a) \in \Dv_i$ & $-2<M/t<-1$ & $-1<M/t<0$ & $0<M/t<1$ & $1<M/t<2$ 
\\ \hline
$(0,0)$ & $-$ & $-$ & $-$ & $-$ 
\\ \hline
$(\pi/2,0)$, $(0, \pi/2)$, $(-\pi/2,0)$, $(0,-\pi/2)$ & $-$ & $-$ & $-$ & $+$ 
\\ \hline
$(\pi/2,\pi/2)$, $(\pi/2,-\pi/2)$, $(-\pi/2,\pi/2)$, $(-\pi/2,-\pi/2)$, $(0, \pi)$, $(\pi, 0)$ & $-$ & $-$ & $+$ & $+$ 
\\ \hline
$(\pi/2,\pi)$, $(\pi, \pi/2)$, $(-\pi/2,\pi)$, $(\pi,-\pi/2)$ & $-$ & $+$ & $+$ & $+$ 
\\ \hline
$(\pi,\pi)$ & $+$ & $+$ & $+$ & $+$ 
\\ \hline
Chern number, $C$ & $1$ & $-3$ & $3$ & $-1$ 
\\ \hline
\end{tabular}
\caption{The sign of $d_z = M - t \cos k_x - t \cos k_y$ at different Dirac points for various Chern insulator phases, along with the Chern number $C = \frac{1}{2} \sum_{\kv \in \Dv_i} {{\rm Sgn}}(\partial_{k_x}\dv\times\partial_{k_y} \dv)_z {{\rm Sgn}}(d_z)$.}
\label{table3}
\end{table}
The lattice models we have introduced apply to the case where the conductivity disperses in frequency but not wave-vector. A more thorough representation of the conductivity tensor which takes both frequency and wave-vector dispersion into account is studied for graphene-like materials in Refs.~\cite{fialkovsky2011} and \cite{rodriguez-lopez2018}, which will be useful in investigating nonlocal contributions to the Casimir force between plasmonic two-dimensional materials.

\subsection{Casimir repulsion between Chern insulators}
\label{sec:chern}

The Casimir-Lifshitz interaction between a pair of Chern insulators of Chern numbers $|C|= 1, 2$ was investigated in Ref.~\cite{rodriguez-lopez2014} for the case of zero temperature. They found that if the plates have opposite signs for the Chern numbers \bing{and the Chern numbers are smaller than 1,} then the Casimir-Lifshitz energy (\ref{lif}) exhibits a peak at an intermediate distance $d_{max} \sim (t_1 t_2)^{1/2} (|C_1 C_2| u_1 u_2)^{-1/2}$ (where $t_i$, $C_i$ and $u_i$ are respectively the hopping parameter, Chern number and mass gap of the $i$th Chern insulator), which is of the order of 200 nm. The peak can be understood as arising from the competition between attraction at short distances and repulsion at sufficiently large distances. 
Reference~\cite{rodriguez-lopez2016} subsequently extends the investigation  to two-dimensional graphene-like materials which can exhibit other topological electronic phases in addition to the quantum anomalous Hall effect (e.g., quantum spin Hall insulator, spin valley polarized semimetal, etc) upon tuning certain external parameters (such as the intensity of an incident circularly polarized laser light and the strength of an electric field applied perpendicularly to the surface). 

At large distances, the Casimir-Lifshitz energy is dominated by the static limit of the conductivity, where the longitudinal conductivity vanishes and the Hall conductivity is quantized in units of $e^2/h$ according to the Chern number. 
The energy per unit area then becomes~\cite{marachevsky2017}
\be
E(d) = - \frac{\hbar c}{8 \pi^2 d^3} \, {{\rm Re}} \, {{\rm Li}}_4 
\left(  
\frac{C_1 C_2 \alpha^2}{(C_1 \alpha + i)(C_2 \alpha + i)}
\right). 
\label{eq:Li}
\ee
In the limit that $|C|\alpha \to \infty$, the above becomes equal to the Casimir-Lifshitz energy per unit area for a pair of perfect conductors, i.e., $E(d) \to - \hbar c \pi^2/(720 d^3)$, and the force is always attractive, independent of the sign of $C_1$ and $C_2$~\cite{footnote:bigC}. 

For the case where \bing{$|C|\alpha \ll 1$}, the energy per unit area can be expanded as a series in powers of $C\alpha$; the leading order term is given by  
\be
E(d) \approx \frac{\hbar c \alpha^2 C_1 C_2}{8\pi^2 d^3},  
\label{ECI}
\ee
where $\alpha$ is the fine structure constant. 
Here, we have a plus sign in the energy, instead of a minus which was obtained in Ref.~\cite{rodriguez-lopez2014}. 
The minus sign in Ref.~\cite{rodriguez-lopez2014} can be traced to the use of the Lifshitz formula (\ref{lif}) with $\Rv_1$ instead of $\Rv_1'$~\cite{fialkovsky2018}. As we show in App.~\ref{app:1}, the off-diagonal entries in $\Rv'$ differ from the corresponding ones in $\Rv$ by a sign factor, this implies that in the large separation regime, the sign in $E(d)$ is positive rather than negative. 

Thus from the energy expression \bing{Eq.~(\ref{ECI})}, we see that the Casimir-Lifshitz force can become repulsive at sufficiently large distances for Chern insulators with Chern numbers of the same sign, and furthermore, the strength of the repulsion tends to increase with an increase of the Chern number. 
This also corresponds to a configuration of Chern insulators with Hall current axial vectors pointing in the same direction, as the sign of the Chern number is related to the sign of $\sigma_{xy}$. 
Conversely, for Chern numbers of opposite signs (a configuration corresponding to Hall current axial vectors pointing in opposite directions), the force is always attractive. 
The sign of the Chern number (and correspondingly, also the Casimir force) can be changed simply by flipping over the Chern insulator. 
\bing{For $C \equiv C_1=C_2$, the possibility of repulsion is determined by condition $|C|\alpha \lesssim 1.032502$; the force actually reverts to attraction for $|C| \gtrsim 1.032502/ \alpha $~\cite{marachevsky2019}. Here, the number $1.032502$ follows from Eq.~(\ref{eq:Li}) for $C \equiv C_1=C_2$.}    

At smaller distances, contributions at all (imaginary) frequencies have to be taken into account. The longitudinal conductivity decays more slowly than the Hall conductivity as a function of imaginary frequency, and is thus dominant at smaller distances. In the short distance limit, where high frequency modes dominate the contribution to the Casimir-Lifshitz energy, and assuming $|C| \alpha < 1$, the longitudinal (Hall) conductivity is found to decay with frequency as $\alpha s_{xx}/\omega$ ($C \alpha s_{xy}/\omega^2$)~\cite{rodriguez-lopez2014}. The longitudinal conductivity then dominates, which has the overall effect in making the Casimir-Lifshitz force attractive at such distances. The short-distance limit of (\ref{lif}) is given by 
\be
E(d) \approx - \frac{3\hbar c \sqrt{\alpha}}{128 d^{5/2}} 
\frac{ \sqrt{ s_{xx,1} s_{xx,2} } }
{\sqrt{s_{xx,1}} + \sqrt{s_{xx,2}} }, 
\ee
i.e., the force is attractive and undergoes a fractional power-law decay as $d^{-7/2}$. 

Though the case of nonzero temperature is not discussed in Ref.~\cite{rodriguez-lopez2014}, we may sketch out some qualitative expectations for Chern insulators induced by surface ferromagnetization. 
If the temperature is nonzero, a thermal lengthscale appears, i.e., $\lambda_T = \hbar c/k_BT$. At a given nonzero temperature, there exists a so-called Lifshitz regime~\cite{lifshitz1955}, which corresponds to separation distances $d$ larger than $\lambda_T$. In this regime, the factor $e^{-2q_n d}$ in the Casimir-Lifshitz energy is exponentially suppressed for $n > 0$; thus the energy is dominated by the contribution from the zero Matsubara frequency mode.   
On the other hand, for an arbitrary separation $d$, the factor $e^{-2q_n d}$ is generally not small and the contribution from nonzero Matsubara frequency modes may not be negligible. However, the nonzero Matsubara frequency mode contribution would become negligible if $\xi_1 = 2\pi k_BT/\hbar > E_{gap}/\hbar$, where $E_{gap}$ is the largest value of the band-gap in the insulator; this is because the conductivity response at imaginary frequencies quickly decays away for energies larger than $E_{gap}$. The foregoing inequality also implies that the nonzero Matsubara frequency modes can be ignored if $T > E_{gap}/(2\pi k_B)$. 

Thus we expect the Casimir-Lifshitz energy behavior to depend on the interplay between the transition temperature of the surface ferromagnetism (which leads to the anomalous quantum Hall effect) and the high temperature limit (set by $E_{gap}/k_B$). 
The band-gap in a Chern insulator can be as small as 10 meV for silicene with transition metal adatoms~\cite{zhang2013} or as large as 340 meV for stanene honeycomb lattices in which one sublattice is passivated by halide atoms whilst the other sublattice is not~\cite{wu2014}. For $E_{gap} \sim 10$ meV, the high temperature limit is $18.5$ K, whilst for $E_{gap} \sim 340$ meV, the high temperature limit is much larger at $627$ K. 
The ferromagnetic transition temperature is typically of the order of milli-Kelvins~\cite{chang2013,chang2015} or a few Kelvins~\cite{zhao2020}, though ferromagnetic transitions as high as $T = 243$ K (for passivated stanene) and $T = 509$ K (for passivated germanene) have also been predicted~\cite{wu2014}. 
If the ferromagnetic transition temperature $T_f$ is larger than $E_{gap}/(2\pi k_B)$, we can imagine setting the Chern insulators at a temperature larger than $E_{gap}/(2\pi k_B)$ but still lower than $T_f$, so that the Casimir-Lifshitz energy is again dominated by the zero (Matsubara) frequency contribution, for which the Hall conductivity is quantized and the longitudinal conductivity vanishes, and correspondingly, the force can be made repulsive. 
On the other hand, if $T_f < E_{gap}/(2\pi k_B)$, to retain the quantum anomalous Hall effect one needs to work at temperatures smaller than $T_f$, and thus all Matsubara frequency contributions have to be taken into account. As noted earlier, this tends to have the effect of making the force attractive rather than repulsive, because of the longitudinal conductivity tends to dominate over the Hall conductivity in the broadband frequency regime. 

Reference~\cite{muniz2021} explores the Casimir effect of a pair of coplanar Chern insulators induced by the application of an electric field and irradiation with a circularly polarized laser, while being also subjected to strong perpendicular magnetic fields. The imposed magnetic field gives rise to a quantum Hall effect and an additional Chern number associated with Landau levels. They find that the presence of three external control parameters (electric field strength, laser light intensity and chemical potential) allows for better control of the magnitude and sign of the Casimir force. In particular, they examine the dissipationless and dissipative cases at nonzero temperature. In the Lifshitz regime, the dissipationless case leads to a Casimir-Lifshitz energy per unit are given asymptotically by $E = \pi \sigma_{xy,1} \sigma_{xy,2} k_B T/(2d^2)$; the Hall conductivity is proportional to the sum of a Chern number induced by electric field and laser light irradiation and another Chern number induced by the quantum Hall effect, and the Casimir force can be repulsive. On the other hand, for sufficiently strong dissipation, the Casimir-Lifshitz energy per unit area becomes asymptotically given by $E = -\zeta_R(3) k_B T/(16\pi d^2)$ (where $\zeta_R(3) \approx 1.202$), and the Casimir force is attractive.

\subsection{nondispersive Chern insulator: Chern-Simons surface}
\label{sec:CS}

There have been a sequence of studies of the Casimir effect for surfaces characterized by the Chern-Simons electromagnetic action~\cite{bordag2000,markov2006,marachevsky2010,marachevsky2017,buhmann2018,marachevsky2019}, and a repulsive Casimir force is also predicted~\cite{marachevsky2017,marachevsky2019}. 
These surfaces can be identified with the limit of nondispersive conductivity in the Chern insulator, which we show in what follows. 
An intimation that the Chern-Simons surface is nondispersive is given by the fact that it is described by an electromagnetic action; 
generally, dispersion cannot be captured by an action~\cite{LL8}. 
 
We write the Chern-Simons action~\cite{marachevsky2019}:
\be
S = - \frac{1}{4} \int\! dt \, d^3x \, F_{\mu\nu} F^{\mu\nu} + \frac{a}{2} \int\! dt \, d^2x \, \epsilon^{3\sigma\mu\nu} A_\sigma F_{\mu\nu}
\ee
Here, the spacetime indices $\mu, \nu, \sigma = 0,1,2,3$, and $0,1,2$ and 3 refer respectively to the time direction $t$ and the spatial directions $x$, $y$ and $z$, $d^3 x \equiv dx\,dy\,dz$, $d^2x \equiv dx\,dy$, and $\epsilon^{\rho\sigma\mu\nu}$ is the completely antisymmetric tensor with cyclic (anticyclic) permutations of $\epsilon^{0123}$ being equal to 1 $(-1)$. 
The above action is expressed in the language of four-vectors~\cite{LL2}, where the electromagnetic field tensor is given by $F_{\mu\nu} = \partial_\mu A_\nu - \partial_\nu A_\mu$, the four-gradient by $\partial_\mu \equiv ((1/c)\partial/\partial t, \bm{\nabla})$ (where $\bm{\nabla} = (\partial/\partial x, \partial/\partial y, \partial/\partial z)$ is the three-dimensional gradient operator), and the four-potential by $A_\mu \equiv (\phi, -\Av)$ (where $\Av$ is the three-dimensional vector potential and $\phi$ is the scalar potential). The four-gradient and four-potential with raised indices are given by $\partial^\mu = ((1/c)\partial/\partial t, - \bm{\nabla})$ and $A^\mu = (\phi, \Av)$. 

By taking the variation of $S$, we obtain 
\ba
\delta S 
&\!\!=\!\!& 
- \frac{1}{2} \int\! dt\,d^3x \, F^{\mu\nu} \delta F_{\mu\nu} 
+ \frac{a}{2} \int\! dt\,d^2x \, \epsilon^{3\sigma\mu\nu} A_\sigma \delta F_{\mu\nu} 
+ \frac{a}{2} \int\! dt\,d^2x \, \epsilon^{3\sigma\mu\nu} \delta A_\sigma F_{\mu\nu} 
\nonumber\\
&\!\!=\!\!& 
- \int\! dt\,d^3x \, F^{\mu\nu} \partial_\mu \delta A_\nu 
+ a \! \int\! dt\,d^2x \, \epsilon^{3\sigma\mu\nu} A_\sigma \partial_\mu \delta A_\nu 
+ a \! \int\! dt\,d^2x \, \epsilon^{3\sigma\mu\nu} \delta A_\sigma \partial_\mu A_\nu 
\nonumber\\
&\!\!=\!\!& 
\int\! dt\,d^3x \, \delta A_\nu \partial_\mu F^{\mu\nu} 
+ 2a \! \int\! dt\,d^2x \, \epsilon^{3\sigma\mu\nu} \delta A_\sigma \partial_\mu A_\nu, 
\nonumber 
\ea
where on going to the second equality, we have made use of the antisymmetry properties $F_{\mu\nu} = - F_{\nu\mu}$ and $\epsilon^{3\sigma\mu\nu} = - \epsilon^{3\sigma\nu\mu}$, and on going to the third equality we have integrated by parts. The variational principle $\delta S = 0$ leads immediately to 
\be
\label{eom}
\partial_\mu F^{\mu\sigma} + 2a \, \delta(z) \epsilon^{3\sigma\mu\nu} \partial_\mu A_\nu = 0. 
\ee
The electric and magnetic fields are given by $\Ev = -\bm{\nabla}\phi - (1/c) (\partial\Av/\partial t)$ and $\Bv = \bm{\nabla} \times \Av$. As $\bm{\nabla} \times \Bv = \bm{\nabla} \times (\bm{\nabla} \times \Av) = \bm{\nabla} (\bm{\nabla} \cdot \Av) - \nabla^2 \Av$, $\partial_\mu \partial^\mu = (1/c^2) (\partial^2/\partial t^2) - \nabla^2$ and $\partial_\mu A^\mu = (1/c) (\partial\phi/\partial t) + \bm{\nabla} \cdot \Av$, we find that taking the $\sigma = 1$ component of Eq.~(\ref{eom}) leads to 
\ba
&&\partial_\mu F^{\mu 1} + 2a \, \delta(z) (\epsilon^{3102} \partial_0 A_2 + \epsilon^{3120} \partial_2 A_0) 
\nonumber\\
&\!\!=\!\!& 
\frac{1}{c^2} \frac{\partial^2 A_x}{\partial t^2} 
- \nabla^2 A_x 
+ \frac{1}{c} \frac{\partial^2\phi}{\partial t \partial x} 
+ \frac{\partial}{\partial x} \bm{\nabla} \!\cdot\! \Av 
- 2a \, \delta(z) 
\bigg( 
\frac{1}{c} \frac{\partial A_y}{\partial t} + \frac{\partial \phi}{\partial y} 
\bigg) 
\nonumber\\
&\!\!=\!\!& 
(\bm{\nabla} \!\times\! \Bv)_x - \frac{1}{c} \frac{\partial E_x}{\partial t} 
+ 2a \, \delta(z) E_y = 0, 
\ea
whilst taking the $\sigma = 2$ component of Eq.~(\ref{eom}) leads to 
\ba
&&\partial_\mu F^{\mu 2} + 2a \, \delta(z) (\epsilon^{3210} \partial_1 A_0 + \epsilon^{3201} \partial_0 A_1) 
\nonumber\\
&\!\!=\!\!& 
\frac{1}{c^2} \frac{\partial^2 A_y}{\partial t^2} 
- \nabla^2 A_y 
+ \frac{1}{c} \frac{\partial^2\phi}{\partial t \partial y} 
+ \frac{\partial}{\partial y} \bm{\nabla} \!\cdot\! \Av 
+ 2a \, \delta(z) 
\bigg( 
\frac{1}{c} \frac{\partial A_x}{\partial t} + \frac{\partial \phi}{\partial x} 
\bigg) 
\nonumber\\
&\!\!=\!\!& 
(\bm{\nabla} \!\times\! \Bv)_y - \frac{1}{c} \frac{\partial E_y}{\partial t} 
- 2a \, \delta(z) E_x = 0. 
\ea
Together, the above equations imply 
\begin{subequations}
\ba
(\bm{\nabla} \!\times\! \Bv)_x &\!\!=\!\!&  \frac{1}{c} \frac{\partial E_x}{\partial t} 
- 2a \, \delta(z) E_y,
\\
(\bm{\nabla} \!\times\! \Bv)_y &\!\!=\!\!&  \frac{1}{c} \frac{\partial E_y}{\partial t} 
+ 2a \, \delta(z) E_x.
\ea
\end{subequations}
As we saw earlier in this section, the static surface conductivity tensor for a Chern insulator is given by Eqs.~(\ref{sigma-chern-static}). Comparing with the Maxwell equation, $\bm{\nabla} \!\times\! \Bv = (1/c) (\partial \Ev / \partial t) + (4\pi/c) \, \delta(z) \, \bm{\sigma} \cdot \Ev$, we find that the Chern-Simons model is equivalent to the nondispersive conductivity limit of the Chern insulator model if we equate $a = - (2\pi/c) \sigma_{xy} = - C \alpha$. This equality leads to the reflection coefficients found for the nondispersive Chern insulator, Eqs.~(\ref{r-chern-static}), which are equivalent to the reflection coefficients found in Ref.~\cite{buhmann2018} (cf. Eqs.~(A14) and (A15)).

\section{Relativistic Quantum Hall effect (RQHE) in graphene}

\begin{figure}[h]
    \centering
      \includegraphics[width=0.8\textwidth]{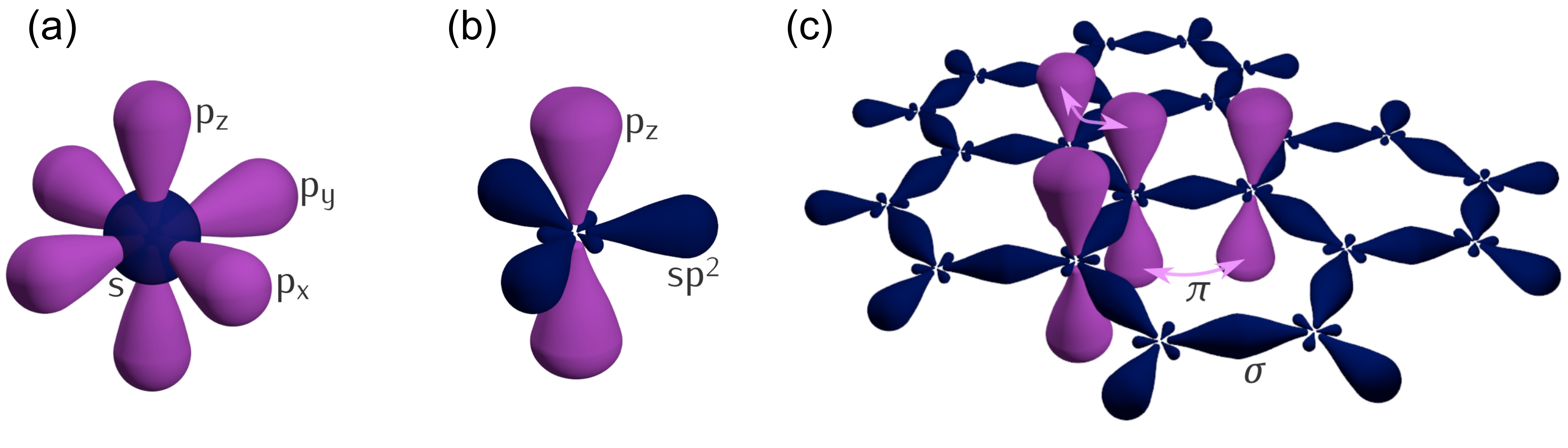}
      \caption{The hybridisation of outermost atomic orbitals in carbon atoms as they form graphene: (a)~orbitals in a free carbon atom, which has 4 outermost electrons; (b)~hybridisation of the 2s, $2{{\rm p}}_x$ and $2{{\rm p}}_y$ orbitals to ${{\rm sp}}^2$ orbitals comprising 3 electrons and a ${{\rm p}}_z$ orbital containing one electron; (c)~the ${{\rm sp}}^2$ orbitals of neighboring atoms overlap to form $\sigma$ bonds in graphene, whilst the ${{\rm p}}_z$ orbitals form $\pi$ bonds. Figure adapted from Wikipedia: \url{https://en.wikipedia.org/wiki/Graphene}} 
      \label{graphene-orbitals}
\end{figure} 
As our next example, we consider the Casimir effect on monolayer graphene subjected to a perpendicular magnetic field. Graphene is a well-explored topic with an extensive literature and has been the subject of very detailed reviews (such as Ref.~\cite{castro-neto2009}). 
A graphene monolayer essentially consists of carbon atoms arranged in a hexagonal pattern in a two-dimensional plane, with the 2s, $2{{\rm p}}_x$ and $2{{\rm p}}_y$ atomic orbitals of each atom hybridising and overlapping with the corresponding ones from neighboring atoms to form $\sigma$ bonds (cf. Fig.~\ref{graphene-orbitals}). Three outermost electrons from each atom are involved in $\sigma$ bonding, whilst the remaining outermost electron resides in the unhybridised $2 {{\rm p}}_z$ orbital which is then free to tunnel (or ``hop'') across to $2 {{\rm p}}_z$ orbitals of neighboring atoms. In the tight-binding approximation, one assumes that the $2 {{\rm p}}_z$ electron from a given atom can at most hop across to nearest neighboring atoms, and the probability of it hopping to next nearest neighbors is much smaller and can be neglected to a first approximation. The resulting tight-binding Hamiltonian exhibits a conduction band and a valence band in wavevector space, these bands are typically separated by a bandgap but they do touch at two special points (inside the Brillouin zone) called the K and K' points. If the Hamiltonian is linearized around each of these points, it has a structure similar to that of a Dirac electron living in two dimensions, i.e., 
\be
H^\xi = \xi \hbar v_F (k_x \sigma_x + k_y \sigma_y), 
\ee
where $\xi=1$ ($\xi=-1$) if the tight-binding Hamiltonian is linearized around the K (K') point, $\sigma_x$ and $\sigma_y$ are Pauli matrices, $v_F = 3ta/2\hbar \approx 10^6 \, {{\rm m.s^{-1}}}$ is the Fermi velocity, $a = 0.24 \, {{\rm nm}}$ is the lattice constant of graphene, and $\kv$ is the wavevector. Here, the use of Pauli matrices is related to the existence of two distinct sublattices in graphene. 

\subsection{Magneto-optical conductivity}

A relativistic quantum Hall effect (RQHE), in which Landau levels appear and the energy of the $n$th Landau level varies with $\sqrt{n|B|}$, emerges if we subject monolayer graphene to a perpendicularly directed magnetic field. The electronic properties of such a system are reviewed in Ref.~\cite{goerbig2011}; thus in what follows, we merely summarise the salient points which we need to obtain the conductivity tensor required for calculating the Casimir force on such a system. 

In practice, many studies of the RQHE in monolayer graphene adopt the approximation of the tight-binding lattice Hamiltonian by the Hamiltonian of a two-dimensional Dirac electron in a magnetic field, 
\be
H_B^\xi = \xi v_F (\Pi_x \sigma_x + \Pi_y \sigma_y), 
\label{HPP}
\ee
where $\bm{\Pi} = \hbar\kv + (e/c) \Av$ is the canonical momentum, and $\Av$ is the vector potential induced by the magnetic field. 
The above approximation is valid in the regime where the so-called magnetic length $\ell_B \equiv \sqrt{\hbar/(e|B|)} \approx 25.68 \, {{\rm nm}}/\sqrt{|B|({{\rm T}})}$ is much greater than $a$. This is the case for magnetic fields in realistic settings, which tend to be smaller than 45 T. 
For such a regime, one can thus construct ``ladder operators" $\hat{a}$ and $\hat{a}^\dagger$ that satisfy the harmonic oscillator commutator relation $[\hat{a}, \hat{a}^\dagger] = 1$: 
\be
\hat{a} = \frac{\ell_B}{\sqrt{2} \hbar} (\Pi_x - i\Pi_y), 
\,\,
\hat{a}^\dagger = \frac{\ell_B}{\sqrt{2} \hbar} (\Pi_x + i\Pi_y). 
\ee
These operators enable one to re-express the Hamiltonian (\ref{HPP}) as 
\be
H_B^\xi = \frac{\xi \sqrt{2} \hbar v_F}{\ell_B} 
\begin{pmatrix}
0 & \hat{a} \\
\hat{a}^\dagger & 0
\end{pmatrix}, 
\ee
which have eigenspinors $|\psi_{\lambda, n}^\xi \rangle$ given by 
\be
H_B^\xi | \psi_{\lambda,n}^\xi \rangle = \varepsilon_{\lambda,n} | \psi_{\lambda,n}^\xi \rangle, 
\ee
where the energy eigenvalue $\varepsilon_{\lambda,n} = \lambda \sqrt{2n} \hbar v_F/\ell_B$. 
Here $n = 0, 1, 2, \ldots$, and $\lambda=1$ ($-1$) refers to electron-like (hole-like) states. 
For $n \neq 0$, the eigenspinor is given by
\be
| \psi_{\lambda,n}^\xi \rangle = \frac{1}{\sqrt{2}} 
\begin{pmatrix}
|n-1\rangle
\\
\xi \lambda |n\rangle
\end{pmatrix},   
\ee
where $|n\rangle = \frac{1}{\sqrt{n!}} (\hat{a}^\dagger)^n |0\rangle$, with $\hat{a}|n\rangle = \sqrt{n}|n-1\rangle$, $\hat{a}^\dagger|n\rangle = \sqrt{n+1}|n+1\rangle$, and $\hat{a}^\dagger \hat{a}|n\rangle = n |n\rangle$. 
The last equation indicates that $n$ is a quantum number which can be used to label the states of the system, with $|n\rangle$ then denoting the $n$th Landau level. 
For $n=0$, the analysis is done separately and yields $\psi_0 = (0, \,\, |0\rangle)^{{\rm T}}$, with zero energy eigenvalue. 

As discussed in Ref.~\cite{goerbig2011}, there is a second quantum number (which we denote by $m$) for monolayer graphene in a magnetic field (besides $n$), this is related semiclassically to the translation invariance of the guiding center of a cyclotron orbit in an electron gas placed under a magnetic field. The sum over $m$ is equal to the degeneracy of a given Landau level, i.e., the maximum number of flux lines threading the surface for a given magnetic field strength: $\sum_m = \mathcal{A}/2\pi\ell_B^2$. The complete quantum state is thus given by $| \psi_{\lambda, n}^\xi \rangle \otimes |m\rangle \otimes |s\rangle$, where $s = \uparrow, \downarrow$ characterizes the electron spin polarization. Here it is assumed that the Zeeman splitting is much smaller than the Landau level spacing, so effectively there is spin degeneracy.  


We now turn to the conductivity tensor for monolayer graphene in a magnetic field. This was obtained in Refs.~\cite{gusynin2006} and \cite{gusynin2007} by calculating the polarization tensor in a quantum field theoretic approach. As an alternative, we shall use the Kubo formula to obtain the same result, working with the basis of states $|\psi_{\lambda, n}^\xi\rangle \otimes |m\rangle \otimes |s\rangle$. For simplicity, we consider the clean limit (zero impurity scattering rate). The current operator is given by $j_x = (e/\hbar) \partial H_B^\xi/\partial k_x = \xi e v_F \sigma_x$ and $j_y = (e/\hbar) \partial H_B^\xi/\partial k_y = \xi e v_F \sigma_y$. 
The Kubo formula gives 
\ba
\sigma_{\mu\nu}(\omega) 
&=& 
-
\lim_{\delta\rightarrow 0} 
\frac{i\hbar e^2 v_F^2}{\mathcal{A}}
\sum_{\xi,s,s'} \sum_{m,m'} \sum_{n,n'} \sum_{\lambda,\lambda'}
\frac{\langle s | \otimes \langle m | \otimes \langle \psi_{\lambda, n}^\xi | \sigma_\mu |\psi_{\lambda', n'}^\xi\rangle \otimes |m'\rangle \otimes |s'\rangle}
{\varepsilon_{\lambda, n} - \varepsilon_{\lambda', n'} + \hbar \omega + i\delta} 
\nonumber\\
&&\quad\qquad\qquad\times
\langle s' | \otimes \langle m' | \otimes \langle \psi_{\lambda', n'}^\xi | \sigma_\nu |\psi_{\lambda, n}^\xi\rangle \otimes |m\rangle \otimes | s \rangle
\frac{f(\varepsilon_{\lambda, n}) - f(\varepsilon_{\lambda', n'})}{\varepsilon_{\lambda, n} - \varepsilon_{\lambda', n'}}
\nonumber\\
&=&
-
\lim_{\delta\rightarrow 0} 
\frac{i\hbar e^2 v_F^2}{\pi\ell_B^2}
\sum_{\xi} \sum_{n,n'} \sum_{\lambda,\lambda'}
\frac{\langle \psi_{\lambda, n}^\xi | \sigma_\mu |\psi_{\lambda', n'}^\xi\rangle 
\langle \psi_{\lambda', n'}^\xi | \sigma_\nu |\psi_{\lambda, n}^\xi\rangle}
{\varepsilon_{\lambda, n} - \varepsilon_{\lambda', n'} + \hbar \omega + i\delta} 
\frac{f(\varepsilon_{\lambda, n}) - f(\varepsilon_{\lambda', n'})}{\varepsilon_{\lambda, n} - \varepsilon_{\lambda', n'}}.    
\label{conduct-graphene}
\ea
Let us calculate the Hall conductivity, $\sigma_{xy}$. 
For a given value of $\xi$, we have 
\ba
&&\frac{\langle \psi_{\lambda, n}^\xi | \sigma_x |\psi_{\lambda', n'}^\xi\rangle 
\langle \psi_{\lambda', n'}^\xi | \sigma_y |\psi_{\lambda, n}^\xi\rangle}
{\varepsilon_{\lambda, n} - \varepsilon_{\lambda', n'} + \hbar \omega + i\delta} 
\frac{f(\varepsilon_{\lambda, n}) - f(\varepsilon_{\lambda', n'})}{\varepsilon_{\lambda, n} - \varepsilon_{\lambda', n'}}
\nonumber\\
&=& 
\frac{i}{4} 
\frac{(\delta_{n-1, n'})^2 - (\delta_{n, n'-1})^2}
{\varepsilon_{\lambda, n} - \varepsilon_{\lambda', n'} + \hbar \omega + i\delta} 
\frac{f(\varepsilon_{\lambda, n}) - f(\varepsilon_{\lambda', n'})}{\varepsilon_{\lambda, n} - \varepsilon_{\lambda', n'}} 
\nonumber\\
&=& 
\frac{i}{4} 
\frac{(\delta_{n-1, n'})^2 (f(\varepsilon_{\lambda, n}) - f(\varepsilon_{\lambda', n-1}))}
{(\varepsilon_{\lambda, n} - \varepsilon_{\lambda', n-1} + \hbar \omega + i\delta)(\varepsilon_{\lambda, n} - \varepsilon_{\lambda', n-1})} 
+
\frac{i}{4} 
\frac{(\delta_{n', n+1})^2 (f(\varepsilon_{\lambda, n+1}) - f(\varepsilon_{\lambda', n}))}
{(\varepsilon_{\lambda, n+1} - \varepsilon_{\lambda', n} - \hbar \omega - i\delta)(\varepsilon_{\lambda, n+1} - \varepsilon_{\lambda', n})} 
\ea
and thus 
\ba
&&\lim_{\delta\rightarrow 0} \,
\sum_{\lambda, \lambda'} \sum_{n,n'} 
\frac{\langle \psi_{\lambda, n}^\xi | \sigma_x |\psi_{\lambda', n'}^\xi\rangle 
\langle \psi_{\lambda', n'}^\xi | \sigma_y |\psi_{\lambda, n}^\xi\rangle}
{\varepsilon_{\lambda, n} - \varepsilon_{\lambda', n'} + \hbar \omega + i\delta} 
\frac{f(\varepsilon_{\lambda, n}) - f(\varepsilon_{\lambda', n'})}{\varepsilon_{\lambda, n} - \varepsilon_{\lambda', n'}}
= 
\frac{i}{2} 
\sum_{\lambda, \lambda'} 
\sum_{n=0}^{\infty} 
\frac{f(\varepsilon_{\lambda, n+1}) - f(\varepsilon_{\lambda', n})}{(\varepsilon_{\lambda, n+1} - \varepsilon_{\lambda', n})^2 - (\hbar\omega)^2} 
\nonumber\\
&=& 
-\frac{i}{2} 
\sum_{n=0}^{\infty} 
\big(
f(M_n) + f(-M_n) - f(M_{n+1}) - f(-M_{n+1})
\big)
\left(
\frac{1}{(M_{n+1} - M_n)^2 - (\hbar\omega)^2} 
+ 
\frac{1}{(M_{n+1} + M_n)^2 - (\hbar\omega)^2}
\right),
\ea
which is independent of $\xi$. 
Here we follow the notation of Ref.~\cite{gusynin2007} and define $M_n \equiv \sqrt{2n} \hbar v_F/\ell_B$, which is related to $\varepsilon_{\lambda, n}$ via $\varepsilon_{\lambda, n} = \lambda M_n$. (From the relation it also follows that $M_n = \sqrt{n} M_1$.)
Substituting the above into Eq.~(\ref{conduct-graphene}) and summing over $\xi$ gives
\ba
\sigma_{xy}(\omega) 
&=& 
- \frac{e^3 v_F^2 B}{\pi} 
\sum_{n=0}^{\infty} 
\big(
f(M_n) + f(-M_n) - f(M_{n+1}) - f(-M_{n+1})
\big)
\nonumber\\
&&\times 
\left\{
\frac{1}{(M_{n+1} - M_n)^2 - (\hbar\omega)^2} 
+ 
\frac{1}{(M_{n+1} + M_n)^2 - (\hbar\omega)^2}
\right\}. 
\ea
For the longitudinal conductivity, a similar calculation gives 
\ba
\sigma_{xx} (\omega) 
&=& 
- \frac{i e^3 v_F^2 |B| \hbar\omega}{\pi} 
\sum_{n=0}^{\infty} 
\bigg\{
\frac{f(M_n) + f(-M_{n+1}) - f(M_{n+1}) - f(-M_n)}
{((M_n - M_{n+1})^2 - \hbar^2 \omega^2)(M_{n+1} - M_n)} 
\nonumber\\
&&\qquad\qquad\qquad\qquad- 
\frac{f(M_n) + f(M_{n+1}) - f(-M_n) - f(-M_{n+1})}
{((M_n + M_{n+1})^2 - \hbar^2 \omega^2)(M_{n+1} + M_n)}
\bigg\}. 
\ea
In what follows, we include the chemical potential $\mu$ in the Fermi-Dirac distribution, so that $f(\varepsilon) = (\exp(\beta(\varepsilon-\mu))+1)^{-1} = (1-\tanh(\beta(M_n-\mu)/2))/2$. As mentioned before, $\mu$ can be set by the gating voltage. For a given nonzero $\mu$, one can tune the number of occupied Landau levels by tuning $B$. As one increases $B$, the energy of each Landau level rises and the highest occupied levels become depopulated as they are tuned above the chemical potential. 

A feature of graphene's magneto-optical conductivity is that varying $B$ has the effect of causing the conductivity to vary in a discontinuous manner. 
The discontinuous variation is clearly illustrated for temperatures close to zero. 
At such temperatures, the Fermi-Dirac distribution becomes a Heaviside step function, thus the number of occupied Landau levels changes discontinuously as $B$ is tuned continuously. For simplicity, let us consider the case of zero frequency. The longitudinal conductivity $\sigma_{xx}$ vanishes and the Hall conductivity becomes 
\ba
&&\sigma_{xy} (\omega=0) 
\nonumber\\
&=&
-\frac{2 e^3 v_F^2 B}{\pi} 
\sum_{n=0}^{\infty} 
\big( f(M_n) + f(-M_n) - f(M_{n+1}) - f(-M_{n+1}) \big) 
\frac{2n+1}{M_1^2}
\nonumber\\
&=&
- \frac{e^3 v_F^2 B}{\pi M_1^2} 
\sum_{n=0}^{\infty} (2n+1)
\left( 
\tanh \frac{\beta}{2}(\mu - M_n) + \tanh \frac{\beta}{2}(\mu + M_n) 
- \tanh \frac{\beta}{2}(\mu - M_{n+1}) - \tanh \frac{\beta}{2}(\mu + M_{n+1}) 
\right) 
\nonumber\\
&=&
- \frac{2e^3 v_F^2 B}{\pi M_1^2} 
\left( 
\tanh \frac{\beta}{2} \mu
+
\sum_{n=1}^{\infty}  
\left( 
\tanh \frac{\beta}{2}(\mu - M_n) + \tanh \frac{\beta}{2}(\mu + M_n) 
\right) 
\right).   
\label{hall-current-graphene}
\ea
Here we see that the sign of $B$ controls the sign of the Hall conductivity, so reversing the magnetic field direction causes the Hall current to circulate in the opposite direction. As $T \to 0$, $\tanh (\beta y/2) \to {{\rm Sgn}} (y)$. For $\mu > 0$ and writing the Heaviside step function $\Theta(y) = (1 + {{\rm Sgn}}(y))/2$, the static Hall conductivity in the zero temperature limit becomes 
\be
\sigma_{xy} (\omega=0) \to - \frac{2e^3 v_F^2 B}{\pi M_1^2} \nu_B 
= - \frac{\alpha c \nu_B}{\pi} {{\rm Sgn}} (B),  
\label{qhe}
\ee
where $\nu_B \equiv 1 + 2 \sum_{n=1}^\infty \Theta(\mu - M_n)$ is proportional to the number of occupied Landau levels. Because $\nu_B$ changes discontinuously as $B$ is tuned, $\sigma_{xy}(\omega=0)$ also varies in a discontinuous fashion, exhibiting a sequence of plateaux, which is a manifestation of the relativistic quantum Hall effect in graphene.

\subsection{Casimir-Polder interaction between rubidium atom and RQHE graphene} 

The discontinuous variation of the conductivity tensor leads to a corresponding variation in the behavior of the Casimir-Polder interaction on a ground-state rubidium atom near monolayer graphene subjected to a magnetic field, as found in Ref.~\cite{cysne2014}. 
For an atom positioned at a height $z$ above a surface, the Casimir-Polder energy can be obtained from second-order perturbation theory by considering the effect of the dipole-radiation coupling on the shift in the energy levels of an atom near a surface; the result is found to be~\cite{gorza2006,wylie1985} 
\be
E_{CP}(z) = - k_BT \, {\sum_{n=0}^{\infty}}' \Tr ({{\bm\alpha}}(i\xi_n)\cdot {{\mathbf G}}^R(z, i\xi_n)),
\ee
where the Matsubara frequency $\xi_n = (2\pi k_B T /\hbar) n$, the prime on the sum means that we multiply the $n=0$ term by a factor of $1/2$, ${{\bm\alpha}}$ is the polarizability tensor of the atom and ${{\mathbf G}}^R$ is the reflection Green tensor which describes the electromagnetic response of the surface. 
As in the case of the Casimir-Lifshitz energy, the zero-temperature Casimir-Polder energy can be recovered by making the replacement $2\pi(k_BT/\hbar){\sum_{n=0}^{\infty}}' \to \int_0^\infty d\xi$. 
For ground-state rubidium, the polarizability tensor is symmetric and isotropic (i.e., ${{\bm \alpha}} = \alpha \mathbb{I}$ where $\mathbb{I}$ is the 3x3 identity matrix), so we can express the Casimir-Polder energy as 
\be
E_{CP}(z) = - k_BT \, {\sum_{n=0}^{\infty}}' \alpha(i\xi_n) ( G_{xx}^R(z, i\xi_n) + G_{yy}^R(z, i\xi_n) + G_{zz}^R(z, i\xi_n) ). 
\ee
For a two-dimensional material with surface conductivity, the reflection Green tensor is also known (see e.g. Ref.~\cite{lu2020}): 
\begin{subequations}
\label{dyadic}
\ba
\label{Gxx}
&&G_{xx}^R(z, \omega) = G_{yy}^R(z, \omega) 
=
\frac{i}{2} \int_0^\infty \!\!\! dk_\perp \frac{k_\perp}{k_z}
\Big( 
\Big( \frac{\omega}{c} \Big)^2 r_{ss} - k_z^2 r_{pp} 
\Big)
e^{2i k_z z},
\nonumber\\
&&G_{zz}^R(z, \omega) = 
i \int_0^\infty \!\!\! dk_\perp \frac{k_\perp^3}{k_z} r_{pp} e^{2i k_z z}, 
\label{Gzz}
\ea
\end{subequations}
where $k_\perp = (k_x^2 + k_y^2)^{1/2}$ and $k_z = ((\omega/c)^2 - k_\perp^2)^{1/2}$. 
Writing $k_z(i\xi_n) = i q_n$, i.e., $q_n = ((\xi_n/c)^2+k_\perp^2)^{1/2}$, the Casimir-Polder energy becomes~\cite{bordag-book}
\be
E_{CP} = - k_BT \, {\sum_{n=0}^{\infty}}' \alpha(i\xi_n) 
\int_0^\infty \!\!\! k_\perp dk_\perp 
q_n e^{- 2 q_n z} 
\left( 
2r_p(i\xi_n, k_\perp) 
- \frac{\xi_n^2}{q_n^2 c^2} 
\left(
r_p(i\xi_n, k_\perp) + r_s(i\xi_n, k_\perp)
\right)
\right)
.   
\ee
Here, the reflection coefficients are given by 
\ba
r_{ss}(i\xi_n, k_\perp) &\!\!=\!\!& 
-\frac{(\tsigma_{xx}(i\xi_n))^2 + (\tsigma_{xy}(i\xi_n))^2
+ \frac{\xi_n}{c q_n} \tsigma_{xx}(i\xi_n)}
{(\tsigma_{xx}(i\xi_n))^2 + (\tsigma_{xy}(i\xi_n))^2 + \big( \frac{\xi_n}{c q_n} + \frac{c q_n}{\xi_n} \big) \tsigma_{xx}(i\xi_n) + 1},
\nonumber\\
r_{pp}(i\xi_n, k_\perp) &\!\!=\!\!& 
\frac{(\tsigma_{xx}(i\xi_n))^2 + (\tsigma_{xy}(i\xi_n))^2 
+ \frac{c q_n}{\xi_n} \tsigma_{xx}(i\xi_n)}{(\tsigma_{xx}(i\xi_n))^2 + (\tsigma_{xy}(i\xi_n))^2 + \big( \frac{\xi_n}{c q_n} + \frac{c q_n}{\xi_n} \big) \tsigma_{xx}(i\xi_n) + 1}, 
\ea
where we have defined the dimensionless conductivity tensor $\tsigma_{\mu\nu}(i\xi_n) \equiv (2\pi/c) \sigma_{\mu\nu}(i\xi_n)$. 

In Ref.~\cite{cysne2014}, the behavior of the Casimir-Polder energy $E_{CP}$ in ground-state rubidium atom near a quantum Hall graphene sheet is studied over a range of separations from 50 nm to 20 $\mu$m  for $T = 4 \, {{\rm K}}$ and chemical potential $\mu_c = 0.115 \, {{\rm eV}}$. The authors find that the ratio $E_{CP}(B)/E_{CP}(B=0)$ undergoes a sequence of discontinuous drops as $B$ is increased (with each drop larger than the preceding one), and there exists a critical value $B_c$ of the magnetic field beyond which the behavior of the Casimir-Polder energy remains more or less independent of the value of $B$. The sequence of discontinuous drops can be attributed to the observation that as $B$ is increased, the energy of the highest occupied Landau level crosses from larger to smaller values of $\sqrt{n} M_1$. The difference $(\sqrt{n} - \sqrt{n-1})M_1$ is larger than $(\sqrt{n+1} - \sqrt{n})M_1$, thus each drop is larger than the previous one. For sufficiently large $B$ only the $n=0$ Landau level is occupied, there are no more crossings and thus the Casimir-Polder energy becomes effectively independent of $B$. It is found that the application of the magnetic field can lead to a reduction of as much as $80\%$ in the Casimir-Polder energy, relative to the value in the absence of the field. 

Finally, at sufficiently large distances (i.e., larger than 1 $\mu$m ), the Casimir-Polder energy exhibits a sequence of plateaux as $B$ is varied. This is related to the fact that in the large distance limit, 
the conductivity tensor becomes well-approximated by its static limit, where the longitudinal conductivity vanishes and only the static Hall conductivity contributes. 
As we have seen in Eq.~(\ref{qhe}), the static Hall conductivity is characterised by a similar sequence of plateaux as $B$ is varied.  
Conversely, as the temperature is increased, the decrease of $E_{CP}$ with the increase of $B$ eventually becomes continuous. 
This is related to the disappearance of the quantum Hall effect at higher temperatures.

\subsection{Casimir interaction between two RQHE graphene monolayers}

One can also investigate the Casimir force between two planar graphene monolayers. 
Here we provide a quick summary of some important results for the case of interacting graphene layers in the absence of a magnetic field, before moving on to consider the topologically nontrivial case where a magnetic field is present. The case without a magnetic field has been investigated by many workers in the past~\cite{bordag2012,klimchitskaya2014a,klimchitskaya2014b,fialkovsky2011,khusnutdinov2015,khusnutdinov2018,drosdoff2010,gomez-santos2009}. At zero temperature, if one takes into account that the longitudinal conductivity of graphene is approximately constant over a wide range of frequencies with a value $\sigma_{xx} = e^2/(4\hbar)$~\cite{gusynin2007,falkovsky2007,nair2008} and consider the limit $2\pi \sigma_{xx}/c \ll 1$, the corresponding Casimir-Lifshitz energy between two graphene monolayers can be approximated by $E(d) \approx -\hbar c \alpha/(32\pi d^3)$~\cite{drosdoff2010}, i.e., the energy is linear in the fine structure constant.  
At finite temperature, plasmons are thermally excited. Because of the gaplessness of graphene, there is a zero-frequency mode plasmon. 
\bing{According to an argument in Ref.~\cite{gomez-santos2009}, the thermal lengthscale at which the Lifshitz regime begins is shortened from $\lambda_T = \hbar c/k_B T$ to $\lambda_{T,p} = \hbar v_F/k_B T$ (where $v_F$ is the Fermi velocity, the velocity at which the plasmons propagate), as there is now also a zero frequency plasmon mode whose correlations give rise to the thermal Casimir force, $E(d) \approx -\zeta(3) k_B T/(16\pi d^2)$. 
Retardation effects do not modify the power-law scaling.  The Lifshitz regime can thus occur at the much shorter separation of 150 nm rather than 7 $\mu$m  at room temperature, and there can be a giant thermal effect, i.e., the thermal Casimir force is much larger. 
On the other hand, calculations based on the polarization tensor evaluated at finite temperature indicate that the thermal lengthscale is mainly determined by the fine structure constant $\alpha$ rather than $v_F$~\cite{fialkovsky2011,klimchitskaya2014a}. 
The giant thermal effect has not been measured, though an experimental proposal has been put forward for detecting it~\cite{bimonte2017}.}  

We now turn to consider the case of graphene monolayers subjected to a perpendicular magnetic field. 
Essentially, the expressions for the conductivity tensor and reflection coefficients remain the same as those that we saw for the Casimir-Polder problem in the previous subsection, the only change is that one works with the Casimir-Lifshitz energy instead of the Casimir-Polder energy. 
The case of zero temperature is studied in Ref.~\cite{tse2012}. Here we flesh out some details of the calculation for the large separation regime, which highlights the differences and similarities between the Casimir force for two RQHE graphene monolayers and that for two perfect metals. 

\begin{figure}[h]
    \centering
      \includegraphics[width=0.6\textwidth]{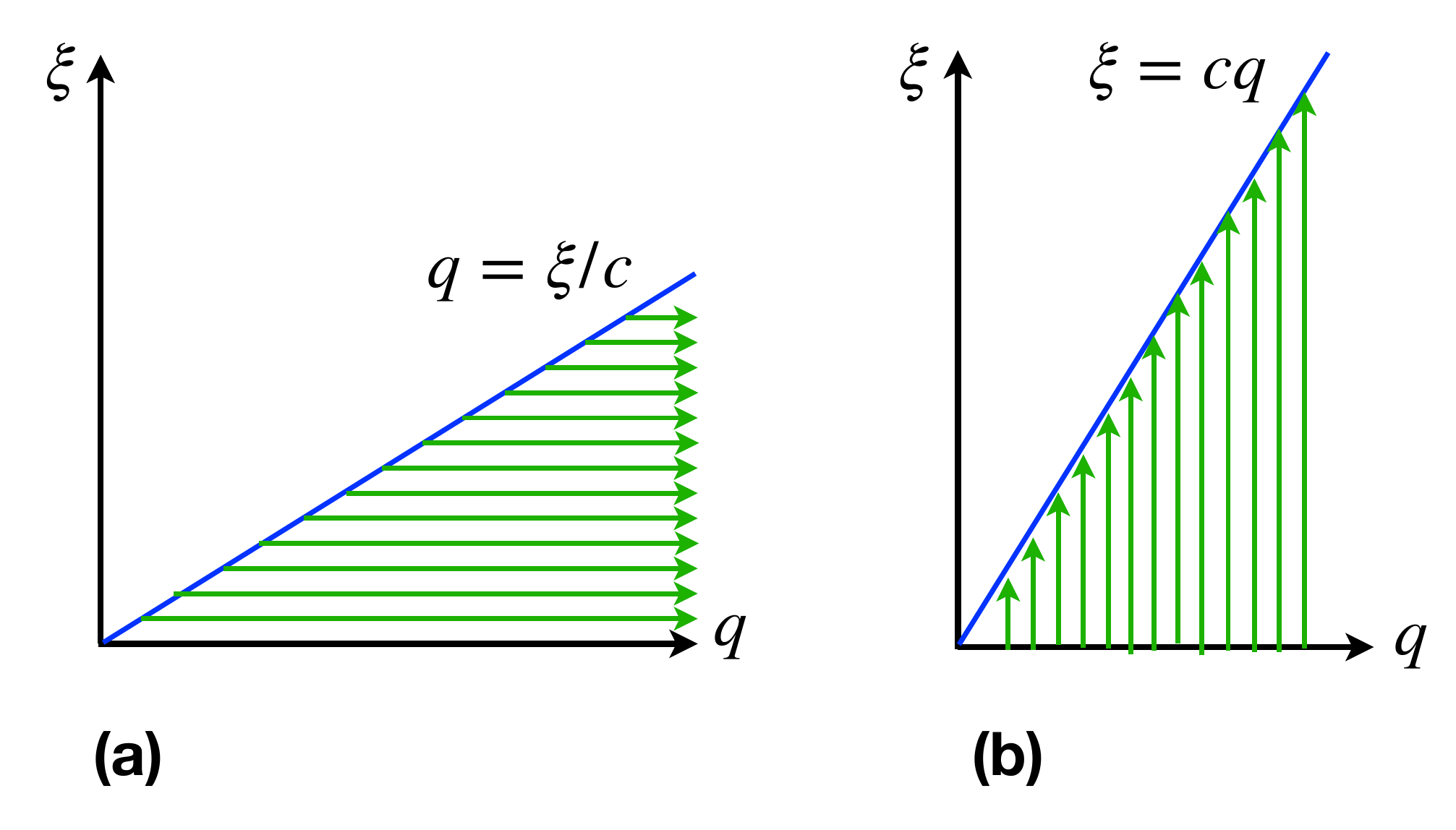}
      \caption{Integration order: (a)~first with respect to $q$, then with respect to $\xi$; (b)~first with respect to $\xi$, then with respect to $q$.} 
      \label{integration-order}
\end{figure} 
In the large separation regime, the exponential factor in the Casimir-Lifshitz energy per unit area (\ref{lif}) is small and so we can expand to leading order, 
\ba
E(d) &=& \frac{\hbar}{2\pi} \int_0^\infty \!\!\! d\xi \int \frac{d^2\kv_\perp}{(2\pi)^2} \, 
\Tr \ln \left( \mathbb{I} - \Rv_1' \cdot \Rv_2 \, e^{-2 q d} \right) 
\nonumber\\
&\approx& 
- \frac{\hbar}{4\pi^2} \int_0^\infty \!\!\! d\xi \int_0^\infty \!\!\! dk_\perp k_\perp  \,  
\Tr \left( \Rv_1' \cdot \Rv_2 \right)
 e^{-2 q d} 
\nonumber\\
&=& 
- \frac{\hbar}{4\pi^2} \int_0^\infty \!\!\! d\xi \int_{\xi/c}^\infty \!\!\! dq \, q  \,  
\Tr \left( \Rv_1' \cdot \Rv_2 \right)
 e^{-2 q d} 
 \nonumber\\
&=& 
- \frac{\hbar}{4\pi^2} 
\int_0^\infty \!\!\! dq \, q  
\int_0^{cq} \!\!\! d\xi \,  
\Tr \left( \Rv_1' \cdot \Rv_2 \right)
 e^{-2 q d}, 
 \label{Ed63}
\ea
where the subscripts 1 and 2 refer to the left and right monolayers, and in the final equality, we have interchanged the order of integrations (cf. Fig.~\ref{integration-order}). In the large separation regime, the reflection coefficients are approximately static, analogous to the case of the Chern insulator. Similar to Eqs.~(\ref{r-chern-static}), the static reflection coefficients for RQHE graphene are given by 
\ba
&&r_{ss}(\omega=0) = - r_{pp}(\omega=0) = - \frac{(\tsigma_{xy}(0))^2}{1+(\tsigma_{xy}(0))^2}, 
\nonumber\\
&&r_{sp}(\omega=0) = r_{ps}(\omega=0) = \frac{\tsigma_{xy}(0)}{1+(\tsigma_{xy}(0))^2}. 
\ea
From Eq.~(\ref{qhe}), $\tsigma_{xy}(0)$ is proportional to the fine structure constant $\alpha$. Thus, to leading order in $\alpha$, $\Tr \left( \Rv_1' \cdot \Rv_2 \right) \approx - 2 \tsigma_{xy,1}(0) \tsigma_{xy,2}(0)$, where the minus sign arises because $r_{ps}' = r_{sp}' = - r_{ps} = - r_{sp}$ (cf. the discussion after Eq.~(\ref{B17})). 
At leading order in $\alpha$, we can approximate the Casimir energy per unit area by
\be
E(d) 
\approx 
\frac{\hbar c \, \tsigma_{xy,1}(0) \tsigma_{xy,2}(0)}{8 \pi^2 d^3}
= 
\frac{\hbar c \alpha^2}{2 \pi^2 d^3} \nu_{B,1} \nu_{B,2} \, {{\rm Sgn}}(B_1 \cdot B_2), 
\ee
where $\nu_{B,i}$ is proportional to the number of occupied Landau levels in the $i$th monolayer. 
The sign of the Casimir force is thus controlled by the relative direction of the magnetic fields acting on the graphene monolayers. As we saw in Eq.~(\ref{hall-current-graphene}), the direction of $B$ dictates the direction of the axial vector associated with the Hall current. If the magnetic fields acting on both monolayers point in the same direction, the Casimir energy is positive and the Casimir force is repulsive at large separations. Conversely, if the magnetic fields are antiparallel, then the Casimir force becomes attractive. This is not unlike the behavior that we have seen for Chern insulators and axionic topological insulators. 

The large-separation Casimir force exhibits a power-law decay of $d^{-4}$, which is  similar to the behavior of the zero-temperature Casimir force between perfect metals, given by $f = - \hbar c \pi^2/240d^4$~\cite{bordag-book}, albeit suppressed by a factor proportional to $\alpha^2$. On the other hand, unlike the case of perfect metals, the Casimir force for RQHE graphene is quantized in units of $3\hbar c \alpha^2/(2\pi^2 d^4)$, as $\nu_{B,i}$ is an integer at zero temperature. The dependence on $\nu_{B,1}\nu_{B,2}$ implies that the Casimir energy between two RQHE graphene monolayers should vary discontinuously as $B$ is varied, and become independent of $B$ if $B$ is sufficiently large. This is not unlike the behavior of the Casimir-Polder force between a rubidium atom and RQHE graphene discussed in the previous subsection. Thus, the large-separation behavior of the Casimir force between a pair of RQHE graphene monolayers subjected to magnetic fields can also provide signatures of the relativistic quantum Hall effect in graphene.

\section{short skit on Weyl semimetals, and Conclusion}
 
Beginning in 2015, in addition to the aforementioned systems, there have also been investigations into the Casimir effect of Weyl semimetals.  
Weyl semimetals~\cite{burkov2014,armitage2018} form another class of topological matter, being three-dimensional solids whose three-dimensional bulk band structure exhibit so-called Weyl nodes, which are points in momentum space at which the valence and conduction bands touch. The Weyl nodes have opposite chiralities, and the dispersion in the immediate vicinity of each node is linear and describable by a Bloch Hamiltonian $H(\kv) = k_x \sigma_x + k_y \sigma_y + k_z \sigma_z$. A weak disorder perturbation has the effect of renormalizing the value of $\kv$, i.e., it can only move the position of the Weyl node in momentum space, but not remove it. Thus, Weyl nodes are topologically robust and do not vanish unless they merge and annihilate each other. 
Weyl semimetals can be either time reversal-symmetric but break inversion symmetry, or inversion-symmetric but break time reversal symmetry. The former do not exhibit the anomalous Hall effect, whereas the latter do. 

For time reversal symmetry-broken Weyl semimetals, it is possible to describe the electromagnetic behavior by an axion field theory 
similar to Eq.~(\ref{S-axion}), but where the axion $\theta$ is a spatially homogeneous parameter for a topological insulator, now $\theta(\rv, t) \equiv 2\bv \cdot \rv - 2 b_0 t$ for a TRSB Weyl semimetal, where $2\bv$ ($2b_0$) is the distance between Weyl nodes in momentum (frequency) space, which can be regarded as chiral gauge fields that couple differently to Weyl fermions of different chiralities. 
The appearance of chiral gauge fields corresponds to the so-called chiral anomaly~\cite{burkov2016}. 
Associated with the time reversal symmetry breaking is the appearance of a three-dimensional bulk Hall conductivity in the Weyl semimetal, $\sigma_{xy} = e^2 b/(2\pi^2\hbar)$~\cite{grushin2012b,chen2013,goswami2013}. 
If the Weyl semimetal has an open surface and the vector $\bv$ is not perpendicular to the open surface, it exhibits surface states known as Fermi arc states. These states are essentially the projections of a two-dimensional Fermi surface onto the open surface. 
Pyrochlore iridates which possess magnetic order have been predicted to be a possible realization of the time reversal symmetry-broken Weyl semimetal~\cite{wan2011}, and such a prediction has been experimentally confirmed~\cite{ueda2018}. 

Thus, similar to the topological systems that we have seen in the foregoing sections, there arises the possibility of Casimir repulsion between time reversal symmetry-broken Weyl semimetals. Such a possibility was first investigated in Ref.~\cite{wilson2015}, for the case of two planar Weyl semimetal slabs of finite thickness which are separated by a vacuum gap. It is assumed that the Weyl cones are not tilted, and the configuration is such that $b_x, b_y, b_0 = 0$ and the open surface is in the xy plane (i.e., $\bv$ is perpendicular to the surface). This configuration implies that there are no Fermi arc states, so the surface is rotationally symmetric. 
By assuming that the diagonal components of the conductivity tensor do not make any contribution, the authors of Ref.~\cite{wilson2015} come to the conclusion 
that the Casimir force between two identical Weyl semimetals can become repulsive at short separations but reverts to attraction at larger separations. 
The investigation was further extended in Ref.~\cite{farias2020} to a pair of semi-infinite, planar Weyl semimetal slabs interacting across a chiral medium. Again by assuming that the diagonal conductivity does not make any contribution, the authors find that if the chiral medium is a Faraday material, the Casimir repulsion can be enhanced, whereas an optically active chiral medium would not modify the Casimir repulsion. Two coplanar Weyl semimetal slabs can also experience a Casimir-Lifshitz torque if the chiral gauge field vectors $\bv_1$ and $\bv_2$ of the first and second slabs are misaligned. This scenario was studied in Ref.~\cite{chen2020} for the case where the Weyl cones are not tilted, $b_0 = 0$, and $\bv_1$ and $\bv_2$ are each parallel with the surface of the slab (so again, there are no Fermi arc states). By disregarding the diagonal conductivity contribution, the authors find that such a chiral anomaly-driven torque can be as strong as the torque between non-topological birefringent media. Lastly, the effects of the diagonal conductivity, Fermi arc states and the tilting of the Weyl cones on the Casimir force are comprehensively studied in Ref.~\cite{rodriguez-lopez2020}. One of the main findings is that taking into account the effect of the diagonal conductivity eliminates the repulsion and results in an \emph{attractive} Casimir force between two Weyl semimetals. 

We have given an overview of the Casimir effect in systems characterized by nonzero topological invariants, in particular, three-dimensional magnetic topological insulators, Chern insulators, graphene sheets subjected to external magnetic fields, and lastly time reversal symmetry-broken Weyl semimetals. A feature common to these systems is the presence of time reversal symmetry-breaking (TRSB) fields, and the TRSB-related quantities (such as the antisymmetric Hall conductivity) also evince their topological origins in being quantized to certain topological invariants in the zero frequency limit, the quantization persisting to an approximate degree in the low frequency regime. The Casimir repulsion is associated with the low frequency behavior of such TRSB-related quantities, and because topological quantization is known to be robust in the presence of weak disorder, the predicted Casimir repulsion has the potential to be similarly robust. 
The sign of the Casimir-Lifshitz force is controlled by the TRSB field, which is easy to tune. 
Thus, the TRSB topological materials are potential candidates for the creation of ``anti-stictive" surfaces. 

More realistically, while the components of NEMS/MEMS could conceivably be coated by a TRSB topological material, the components themselves are typically made of a different material such as gold. There would be a contact potential at the interface which stems from the difference between the work functions of the adjacent materials and also local charge inhomogeneities. To discover the Casimir repulsion in a putative experiment, one would need to account for such electrostatic contributions. Recently, there has indeed been an experiment performed on a topological insulator thin film system (${{\rm Bi_2 Se_3}}$) that takes electrostatic contributions into account~\cite{babamahdi2021}, however the adopted topological insulator was time reversal-symmetric, which excluded the possibility of Casimir repulsion. For a putative experiment to detect Casimir repulsion between Chern insulators, one would also need to account for the effect of the dielectric substrate, as Chern insulators are typically grown by molecular beam epitaxy on such substrates~\cite{chang2013,chang2015}.   
To accurately capture the Casimir force behavior between TRSB topological insulators, one would need to know the frequency dispersion of the axion or Hall conductivity of the materials themselves. Although the frequency dispersion of the axion response in magnetic topological insulators or the conductivity tensor in the quantum anomalous Hall effect (QAHE) can be calculated using lattice models (e.g., via the polarization tensor in five-dimensional QED or the Kubo formula), the experimental determination of such dispersion is much harder due to the low threshold temperature at which the QAHE exists. In practice, the Hall conductivity dispersion for the THz range has been obtained indirectly by measuring the behavior of the Faraday and Kerr rotations~\cite{wu2016}.

\section{acknowledgement} 

The author is grateful to Prof. G. Klimchitskaya for the invitation to contribute the present review. 
He also thanks the three anonymous referees for their constructive comments. 
This work was supported by the Singapore Ministry of Education under Grant No. RG160/19(S) of Tier 1 of the Academic Research Fund. 

\appendix

\section{Kubo formula} 
\label{app:kubo} 

To make our review self-contained, here we explain how we obtain the conductivity formula (\ref{kubo-general}). We follow the approach described in Chap.~1 of Ref.~\cite{czycholl2017}. 
Linear response theory assumes that when a weak time-dependent perturbation field $F(t)$ is switched on, the expectation value of a given physical observable $\hat{B}$ differs from the expectation value in the absence of the perturbation by an amount proportional to the perturbation itself, i.e., 
\be
\langle \hat{B} \rangle_{\rho(t)} - \langle \hat{B} \rangle_{\rho_0} = \int_{-\infty}^{+\infty} dt' F(t') \chi_{\hat{B}, \hat{A}} (t-t'). 
\ee
Here $\langle \ldots \rangle_{\rho_0}$ denotes a Gibbsian statistical average with respect to the equilibrium density operator in the absence of the perturbation, $\langle \ldots \rangle_{\rho(t)}$ denotes a Gibbsian statistical average with respect to the density operator at time $t$ after the perturbation has been switched on, and the perturbed Hamiltonian is given by $H(t) = H_0 - \hat{A} F(t)$, where $H_0$ is the unperturbed Hamiltonian and $\hat{A}$ is an operator that couples the perturbing field to the system. The problem of linear response theory is to determine the susceptibility function $\chi_{\hat{B}, \hat{A}} (t-t')$, and a key result is that this function can be expressed in terms of a general formula involving the unperturbed density operator, viz., 
\be
\chi_{\hat{B}, \hat{A}} (t-t') = \frac{i}{\hbar} \left\langle [ \hat{B}_I(t) , \hat{A}_I(t') ] \right\rangle_{\rho_0} \Theta(t-t'), 
\ee
where the subscript $I$ indicates that the operator in question is in the interaction picture: $\hat{X}_I(t) = e^{iH_0t/\hbar} \hat{X} e^{-iH_0t/\hbar}$, and $\Theta(t)$ is the Heaviside step function equal to 0 (1) for $t < 0$ ($t > 0$). In the frequency domain, 
\ba
\chi_{\hat{B}, \hat{A}} (\omega + i\delta) 
&=& \int_{-\infty}^{+\infty} dt \, \chi_{\hat{B}, \hat{A}}(t) e^{i(\omega + i\delta) t}  
=
 \frac{i}{\hbar} \int_0^{+\infty} \langle [\hat{B}_I(t), \hat{A}] \rangle_{\rho_0} e^{i(\omega+i\delta)t} dt
\nonumber\\
&=& 
\frac{i}{\hbar Z_0} \sum_{p,q} \int_0^\infty dt 
\left(
\langle p | e^{iH_0t/\hbar} \hat{B} e^{-iH_0t/\hbar} | q \rangle \langle q | \hat{A} | p \rangle 
-
\langle p | \hat{A} | q \rangle \langle q | e^{iH_0t/\hbar} \hat{B} e^{-iH_0t/\hbar} | p \rangle 
\right) 
e^{i(\omega+i\delta)t} e^{-\beta E_p}
\nonumber\\
&=&
\frac{i}{\hbar Z_0} \sum_{p,q} \int_0^\infty dt 
\left(
\langle p | \hat{B} | q \rangle \langle q | \hat{A} | p \rangle e^{i\omega_{pq}t}
-
\langle q | \hat{B} | p \rangle \langle p | \hat{A} | q \rangle e^{i\omega_{qp}t}
\right)
e^{i(\omega+i\delta)t} e^{-\beta E_p}
\nonumber\\
&=&
\frac{i}{\hbar Z_0} \sum_{p,q} \int_0^\infty dt 
\langle p | \hat{B} | q \rangle \langle q | \hat{A} | p \rangle e^{i(\omega_{pq}+\omega+i\delta)t}
\big( e^{-\beta E_p} - e^{-\beta E_q} \big) 
\nonumber\\
&=& 
- \frac{1}{\hbar Z_0} \sum_{p,q} \frac{\langle p | \hat{B} | q \rangle \langle q | \hat{A} | p \rangle}{\omega_{pq}+\omega+i\delta} \big( e^{-\beta E_p} - e^{-\beta E_q} \big).
\label{chiBA}
\ea
In the above, $Z_0$ is the partition function, $|p\rangle$ and $|q\rangle$ are many-particle eigenstates of $H_0$, with $E_p$ and $E_q$ their corresponding energy eigenvalues. In the fourth equality we have defined $\hbar\omega_{qp} \equiv E_q - E_p$, and in the fifth equality we have interchanged $q \leftrightarrow p$ in the second term. 

We now turn to obtain the Kubo formula for the conductivity tensor. For the longitudinal conductivity, let us take the perturbation term to be $-x \, \mathcal{E}_x e^{-i(\omega + i\delta)t}$ where $x$ is a position operator and $\mathcal{E}_x e^{-i(\omega + i\delta)t}$ the perturbing electric field, so $\hat{A} = x$ and $F(t) = \mathcal{E}_x e^{-i(\omega + i\delta)t} $, and the perturbing field gives rise to a current $j_x$, which is our observable $\hat{B}$. The presence of $i\delta$ in the exponent ensures that the state at $t=-\infty$ is the unperturbed state. In the absence of the perturbation, the average current vanishes identically: $\langle j_x \rangle_{\rho_0} = 0$. In two dimensions, the current density is given by 
\be
J_x = \frac{1}{\mathcal{A}} \langle j_x \rangle_{\rho(t)} 
= \frac{e}{\mathcal{A}} \chi_{j_x, x}(\omega + i\delta) \mathcal{E}_x e^{-i(\omega + i\delta)t} 
= \sigma_{xx}(\omega) \mathcal{E}_x e^{-i(\omega + i\delta)t}, 
\ee
where $\mathcal{A}$ is the area of the surface over which the current flows. 
We can identify the longitudinal conductivity with the susceptibility $\chi_{j_x, x}$: 
\be
\sigma_{xx}(\omega + i\delta) 
= 
\frac{e}{\mathcal{A}} \chi_{j_x, x}(\omega + i\delta) 
=
- \frac{e}{\mathcal{A} Z_0} \sum_{p, q} \frac{\langle q | j_x |q\rangle \langle q | x |p\rangle}{\hbar\omega + i\delta + E_p - E_q} 
\big( e^{-\beta E_p} - e^{-\beta E_q} \big),  
\ee
where in the second equality we have used Eq.~(\ref{chiBA}). 
We are going to work interchangeably in the first (Schr\"{o}dinger) and second (Fock) quantized pictures whichever is more convenient. In the first quantized picture, the many-particle state $|p\rangle = |\psi_1 \psi_2 \ldots \psi_{N_e} \rangle$, where $\psi_i$ is the wavefunction of the $i$th electron, and $N_e$ is the total number of electrons. In this picture, 
the current operator $j_x = e \, \dot{x} = i (e/\hbar) [H_0, x]$, which implies $\langle q| j_x |p \rangle = i (e/\hbar) \langle q| (H_0 x - x H_0) |p \rangle = i (e/\hbar) (E_q - E_p) \langle q | x | p \rangle$. The longitudinal conductivity is thus given by 
\be
\sigma_{xx}(\omega + i\delta) 
=
- \frac{i\hbar}{\mathcal{A} Z_0}
\sum_{p, q} \frac{\langle p | j_x |q\rangle \langle q | j_x |p\rangle \big( e^{-\beta E_p} - e^{-\beta E_q} \big)}
{(\hbar\omega + i\delta + E_p - E_q) (E_p - E_q)}. 
\label{sigmaxx0}
\ee
We can similarly derive an expression for the Hall conductivity if we take the perturbation term to be $-y \, \mathcal{E}_y e^{-i(\omega + i\delta)t}$, so now $\hat{A} = y$ and $F(t) = \mathcal{E}_y e^{-i(\omega + i\delta)t} $, and the Hall current $j_x$ is our observable $\hat{B}$: 
\be
J_x 
= \frac{1}{\mathcal{A}} \langle j_x \rangle_{\rho(t)} 
= \frac{e}{\mathcal{A}} \chi_{j_x, y}(\omega + i\delta) \mathcal{E}_y e^{-i(\omega + i\delta)t} = \sigma_{xy}(\omega) \mathcal{E}_y e^{-i(\omega + i\delta)t}. 
\ee
By a derivation similar to the one for the longitudinal conductivity, we find 
\be
\sigma_{xy}(\omega + i\delta) 
=
- \frac{i\hbar}{\mathcal{A} Z_0}
\sum_{p, q} \frac{\langle p | j_x |q\rangle \langle q | j_y |p\rangle \big( e^{-\beta E_p} - e^{-\beta E_q} \big)}
{(\hbar\omega + i\delta + E_p - E_q) (E_p - E_q)}. 
\label{sigmaxy0}
\ee
The formulas (\ref{sigmaxx0}) and (\ref{sigmaxy0}) are expressed in terms of many-particle states and energies. To make contact with tight-binding model calculations, we shall make a transformation to single-particle states and energies. In the second quantized or Fock space picture, the many-particle state $|p\rangle = |n_{p_1} n_{p_2} \ldots n_{p_i} \ldots \rangle$, where $p_i$ is the set of quantum numbers used to label an electron in the $i$th single-particle state. For example, for an electron in a solid state system, $p = ( l, \kv, s )$, where $l$ is the band index, $\kv$ is the crystal momentum, and $s$ the electron spin, so $\{ p_i \}$ is then the set of all possible permutations of the collection of numbers $\{ l, \kv, s \}$. 
$n_{p_i}$ is the occupation number (either 0 or 1) for an electron in the single-particle state with the set of quantum numbers $p_i = ( l_i, \kv_i, s_i)$. 
The energy of the many-particle state $|p\rangle$ is thus $E_{p} = \sum_{p_i} n_{p_i} \varepsilon_{p_i}$, where $\varepsilon_{p_i}$ is the energy of an electron occupying the single-particle state $|p_i\rangle$. 

In the Fock space picture, the current operator is given by $\jv = e \sum_{p_i, q_j} \langle p_i | \vv | q_j \rangle c_{p_i}^\dagger c_{q_j}$, where $\vv$ is the velocity operator. 
The conductivity tensor involves the product of matrix elements 
\ba
\label{j-elements}
&&\langle p | j_\mu |q\rangle \langle q | j_\nu |p\rangle
\\
&\propto&
\sum_{p_i, q_l}\sum_{q_j, p_k}
\langle n_{p_1}, n_{p_2}, \ldots, n_{p_i}, \ldots | c_{p_i}^\dagger c_{q_l} | n_{q_1}, n_{q_2}, \ldots, n_{q_l}, \ldots \rangle \langle n_{q_1}, n_{q_2}, \ldots, n_{q_j}, \ldots | c_{q_j}^\dagger c_{p_k} | n_{p_1}, n_{p_2}, \ldots, n_{p_k}, \ldots \rangle.   
\nonumber
\ea
$c_{p_i}^\dagger c_{q_l}$ annihilates an electron with the quantum numbers $q_l$ and creates another electron with the quantum numbers $p_i$, whilst $c_{q_j}^\dagger c_{p_k}$ annihilates an electron with the quantum numbers $p_k$ and creates another electron with the quantum numbers $q_j$. In order for $|q\rangle$ and $\langle q |$ (and likewise for $|p\rangle$ and $\langle p |$) to describe the same many-particle state, we must have $q_l = q_j$ and $p_k = p_i$, so we can replace 
\be
\langle n_{p_1}, \ldots, n_{p_i}, \ldots | c_{p_i}^\dagger c_{q_l} | n_{q_1}, \ldots, n_{q_l}, \ldots \rangle \langle n_{q_1}, \ldots, n_{q_j}, \ldots | c_{q_j}^\dagger c_{p_k} | n_{p_1},  \ldots, n_{p_k}, \ldots \rangle
\to \delta_{q_j, q_l} \delta_{p_i, p_k}
\ee
In order that the current matrix elements in Eq.~(\ref{j-elements}) be nonzero, the single-particle states in $|p\rangle$ and $|q\rangle$ (and their corresponding energies) have to coincide except for $|q_l\rangle (=|q_j\rangle)$ and $|p_k\rangle (=|p_i\rangle)$. This implies that $E_p - E_q = \varepsilon_{p_k} - \varepsilon_{q_j} = \varepsilon_{p_i} - \varepsilon_{q_l}$. 
Thus we can write 
\be
e^{-\beta E_p} - e^{-\beta E_q} 
= 
e^{- \beta \sum_{\ell \neq p_i, q_j} n_\ell \varepsilon_\ell} 
\left( e^{- \beta \varepsilon_{p_i}} - e^{- \beta \varepsilon_{q_j}} \right), 
\ee
where $\ell$ denotes the set of quantum numbers used to label a single-particle state. 
The partition function is
\be
Z_0 = \sum_p e^{-\beta E_p} = \sum_{\{n_\ell \}} e^{-\beta \sum_\ell n_\ell \varepsilon_\ell}
= \sum_{n_1=0,1} \!\!\! e^{-\beta n_1 \varepsilon_1} 
   \sum_{n_2=0,1} \!\!\! e^{-\beta n_2 \varepsilon_2} 
   \ldots 
= \prod_\ell \left( 1 + e^{-\beta \varepsilon_\ell} \right).
\ee
Writing $F(E_p - E_q)$ for an arbitrary function of the energy difference $E_p - E_q$, we have 
\ba
&&\frac{1}{Z_0} \sum_{p, q} 
\langle p | c_{p_i}^\dagger c_{q_l} | q \rangle
\langle q | c_{q_j}^\dagger c_{p_k} | p \rangle 
\big( e^{-\beta E_p} - e^{-\beta E_q} \big)
F(E_p - E_q) 
\nonumber\\
&=&
\frac{\prod_{\ell \neq p_i, q_j} \left( 1 + e^{-\beta \varepsilon_\ell} \right)}
{\prod_{\ell'} \left( 1 + e^{-\beta \varepsilon_{\ell'}} \right)}
\left( e^{- \beta \varepsilon_{p_i}} - e^{- \beta \varepsilon_{q_j}} \right)
F(\varepsilon_{p_i} - \varepsilon_{q_j}) 
\delta_{q_j, q_l} \delta_{p_i, p_k}
\nonumber\\
&=&
\delta_{q_j, q_l} \delta_{p_i, p_k}
\big( f(\varepsilon_{p_i}) - f(\varepsilon_{q_j}) \big)
F(\varepsilon_{p_i} - \varepsilon_{q_j}), 
\ea
where $f(\varepsilon)$ denotes the Fermi-Dirac distribution, 
$f(\varepsilon) = (e^{\beta \varepsilon} + 1)^{-1}$. 
Recalling Eqs.~(\ref{sigmaxx0}) and (\ref{sigmaxy0}), and identifying $F(E_p - E_q)  \to 1/((E_q - E_p) (\hbar\omega + i\delta + E_p - E_q))$, we see that the conductivity tensor can be written as  
\be
\sigma_{\mu\nu}(\omega) = 
-
\frac{i\hbar}{\mathcal{A}}
\sum_{\ell,\ell'} 
\frac{\langle \ell | j_\mu | {\ell'} \rangle \langle {\ell'} | j_\nu | {\ell} \rangle}{\varepsilon_\ell - \varepsilon_{\ell'} + \hbar \omega + i\delta} 
\frac{f(\varepsilon_\ell) - f(\varepsilon_{\ell'})}{\varepsilon_\ell - \varepsilon_{\ell'}}, 
\ee
which is the conductivity formula we use in Eq.~(\ref{kubo-general}), expressed in the basis of single-particle states. 

\section{Fresnel coefficients for a pair of Chern insulators} 
\label{app:1}

\begin{figure}[h]
\centering
  \includegraphics[width=0.7\textwidth]{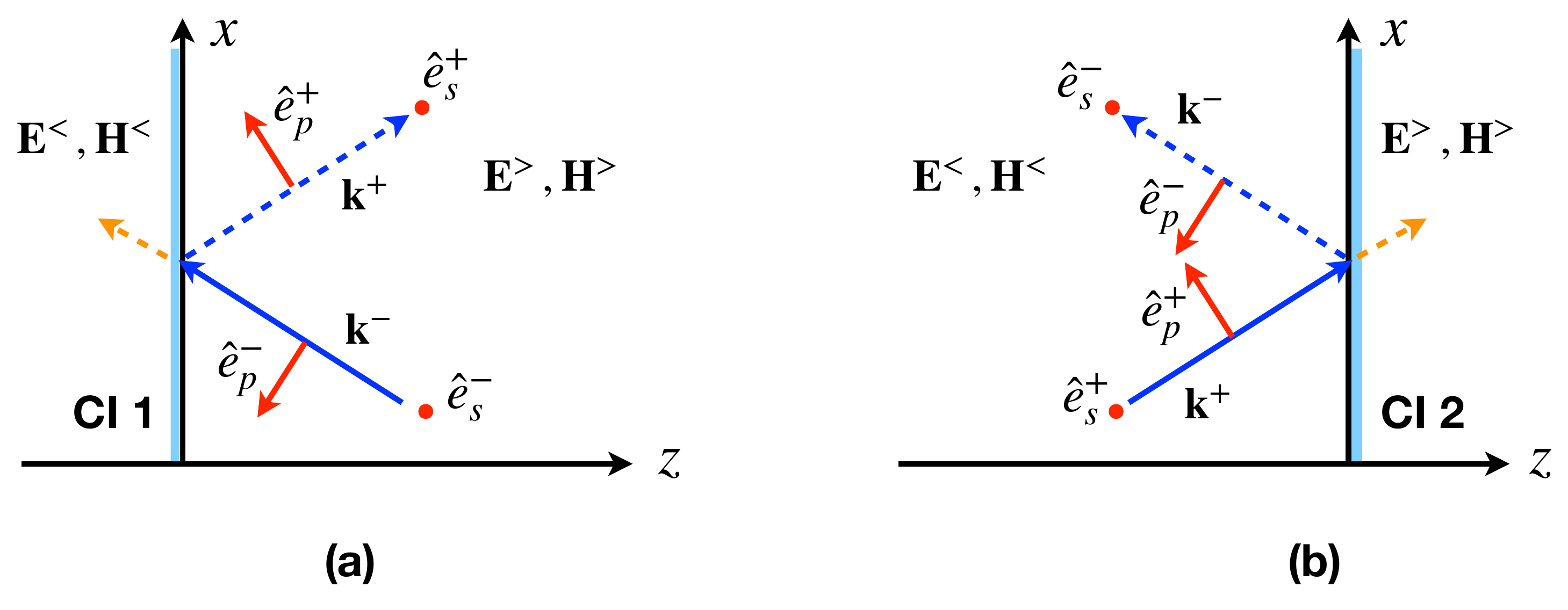}
  \caption{Boundary-value problem for a pair of coplanar Chern insulators (light blue and labeled ``CI 1" and ``CI 2") in vacuum: (a)~ray incident (blue solid arrow) on the left Chern insulator, with wavevector $\kv^- = (k_x, 0, -k_z)^{{\rm T}}$; the transmitted ray (orange dashed arrow) has the same wavevector $\kv_-$ and the reflected ray (blue dashed arrow) has the wavevector $\kv^+ = (k_x, 0, k_z)^{{\rm T}}$. (b)~ray incident on the right Chern insulator, with reflected and transmitted rays. This is a mirror image of the rays in (a). $\Ev^<, \Hv^<$ ($\Ev^>, \Hv^>$) denote the electric and magnetic fields to the left (right) of a Chern insulator. The polarization vectors (red solid arrows) are $\hat{e}_p^\pm = \hat{e}_s^\pm \times \hat{k}^\pm$.} 
  \label{fig:BCs2}
\end{figure}

To obtain the Casimir-Lifshitz energy of a pair of coplanar Chern insulators, we need to derive the reflectivity matrices $\Rv_1'$ and $\Rv_2$ of the left and right Chern insulators (1 and 2 referring to the left and right insulators, respectively). This requires us to solve two boundary value problems, one at each interface (see Fig.~\ref{fig:BCs2}). We shall use the notation $\Ev^<, \Hv^<$ ($\Ev^>, \Hv^>$) to refer to electric and magnetic fields to the left (right) of a given Chern insulator. The boundary conditions are given by~\cite{rodriguez-lopez2014,lu2020} 
\ba
&&E_x^<(z=0) = E_x^>(z=0), \,\, E_y^<(z=0) = E_y^>(z=0); 
\nonumber\\
&&H_x^>(z=0) - H_x^<(z=0) = (4\pi/c) (\sigma_{xx} E_y(z=0) - \sigma_{xy} E_x(z=0)), 
\nonumber\\
&&H_y^<(z=0) - H_y^>(z=0) = (4\pi/c) (\sigma_{xx} E_x(z=0) + \sigma_{xy} E_y(z=0)). 
\label{BCs} 
\ea
The appearance of $\sigma_{xy}$ in the boundary condition equations presupposes a  given orientation of the Hall conductivity. If we apply the boundary condition to each Chern insulators with the same value and sign of $\sigma_{xy}$, we are effectively assuming that the Hall current of each insulator circulates with the same orientation. 

To solve the boundary value problem for a given interface perpendicular to the z direction, let us consider a right- (left-) propagating wavevector by $\kv^\pm = (k_x, 0, \pm k_z)^{{\rm T}}$, and make use of the polarization vectors $\hat{e}_s^\pm = (0,1,0)^{{\rm T}}$ and $\hat{e}_p^\pm = (1/k)(\pm k_z,0,-k_x)^{{\rm T}}$, with $k = |\kv^+| = |\kv^-| = \omega/c$. The polarization vectors and the unit wavevector $\hat{k}^\pm = \kv^\pm/k$ are orthogonal to each other: $\hat{e}_p^\pm = \hat{e}_s^\pm \times \hat{k}^\pm$. 
Furthermore, we have the identities
\begin{subequations}
\ba
&&\hat{x} \cdot \hat{e}_s^\pm = 0, \,\, 
\hat{y} \cdot \hat{e}_s^\pm = 1, \,\, 
\hat{z} \cdot \hat{e}_s^\pm = 0, 
\\
&&\hat{x} \cdot \hat{e}_p^\pm = \pm k_z/k, \,\, 
\hat{y} \cdot \hat{e}_p^\pm = 0, \,\, 
\hat{z} \cdot \hat{e}_p^\pm = -k_x/k
\ea
\end{subequations}
First, consider the boundary value problem for Fig.~\ref{fig:BCs2}(b). 
In the left subspace, we have the incident electric field which can be resolved into independent s and p polarizations, viz., 
\be
\Ev_{{\rm in}}(\rv) 
= 
\big( E_{{\rm in}}^{{\rm (s)}} \hat{e}_s^+ + E_{{\rm in}}^{{\rm (p)}} \hat{e}_p^+ \big) e^{i\kv^+\cdot\rv}
\ee 
and its reflected field,  
\ba
\Ev_{{\rm R}} (\rv)
&=& \big( E_{{\rm R}}^{{\rm (s)}} \hat{e}_s^- + E_{{\rm R}}^{{\rm (p)}} \hat{e}_p^- \big) e^{i\kv^-\cdot\rv}
\nonumber\\
&=& 
\big(
(r_{ss} E_{{\rm in}}^{{\rm (s)}} + r_{ps} E_{{\rm in}}^{{\rm (p)}}) \hat{e}_s^- 
+ 
(r_{sp} E_{{\rm in}}^{{\rm (s)}} + r_{pp} E_{{\rm in}}^{{\rm (p)}}) \hat{e}_p^-
\big) 
e^{i\kv^-\cdot\rv}
\ea
The total electric field is then $\Ev^< = \Ev_{{\rm in}} + \Ev_{{\rm R}}$, i.e., 
\be
\Ev^<(\rv) 
= 
\big( E_{{\rm in}}^{{\rm (s)}} \hat{e}_s^+ + E_{{\rm in}}^{{\rm (p)}} \hat{e}_p^+ \big) e^{i\kv^+\cdot\rv}
+
\big(
(r_{ss} E_{{\rm in}}^{{\rm (s)}} + r_{ps} E_{{\rm in}}^{{\rm (p)}}) \hat{e}_s^- 
+ 
(r_{sp} E_{{\rm in}}^{{\rm (s)}} + r_{pp} E_{{\rm in}}^{{\rm (p)}}) \hat{e}_p^-
\big) 
e^{i\kv^-\cdot\rv}
\label{E<R}
\ee
In the right subspace, we have the transmitted field
\be
\Ev^>(\rv) = 
\big(
(t_{ss} E_{{\rm in}}^{{\rm (s)}} + t_{ps} E_{{\rm in}}^{{\rm (p)}}) \hat{e}_s^+ 
+ 
(t_{sp} E_{{\rm in}}^{{\rm (s)}} + t_{pp} E_{{\rm in}}^{{\rm (p)}}) \hat{e}_p^+
\big) 
e^{i\kv^+\cdot\rv}
\label{E>R}
\ee
The magnetic field is obtained from the Maxwell-Faraday equation, $\Hv = -i(c/\omega) {{\bm \nabla}} \times \Ev$. 
Thus for the left subspace, we have 
\ba
\Hv^< &=& -i(c/\omega) {{\bm \nabla}} \times \Ev^<
\nonumber\\
&=&
\frac{c}{\omega}
\big( E_{{\rm in}}^{{\rm (s)}} \kv^+ \times \hat{e}_s^+ 
+ E_{{\rm in}}^{{\rm (p)}} \kv^+ \times \hat{e}_p^+ \big) e^{i\kv^+\cdot\rv}
\nonumber\\
&&+
\frac{c}{\omega}
\big(
(r_{ss} E_{{\rm in}}^{{\rm (s)}} + r_{ps} E_{{\rm in}}^{{\rm (p)}}) \kv^- \times \hat{e}_s^- 
+ 
(r_{sp} E_{{\rm in}}^{{\rm (s)}} + r_{pp} E_{{\rm in}}^{{\rm (p)}}) \kv^- \times \hat{e}_p^-
\big) 
e^{i\kv^-\cdot\rv}
\nonumber\\
&=&
\frac{ck}{\omega} 
\big(
- E_{{\rm in}}^{{\rm (s)}} \hat{e}_p^+ 
+ E_{{\rm in}}^{{\rm (p)}} \hat{e}_s^+ 
\big) 
e^{i\kv^+\cdot\rv}
+
\frac{ck}{\omega}
\big(
- (r_{ss} E_{{\rm in}}^{{\rm (s)}} + r_{ps} E_{{\rm in}}^{{\rm (p)}}) \hat{e}_p^- 
+ 
(r_{sp} E_{{\rm in}}^{{\rm (s)}} + r_{pp} E_{{\rm in}}^{{\rm (p)}}) \hat{e}_s^-
\big) 
e^{i\kv^-\cdot\rv}
\label{H<R}
\ea
For the magnetic field in the right subspace, we find
\ba
\Hv^> &=& -i(c/\omega) {{\bm \nabla}} \times \Ev^>
\nonumber\\
&=& 
\frac{c}{\omega}
\big(
(t_{ss} E_{{\rm in}}^{{\rm (s)}} + t_{ps} E_{{\rm in}}^{{\rm (p)}}) (\kv^+ \times \hat{e}_s^+) 
+ 
(t_{sp} E_{{\rm in}}^{{\rm (s)}} + t_{pp} E_{{\rm in}}^{{\rm (p)}}) (\kv^+ \times \hat{e}_p^+)
\big) 
e^{i\kv^+\cdot\rv}
\nonumber\\
&=&
\frac{ck}{\omega}
\big(
- (t_{ss} E_{{\rm in}}^{{\rm (s)}} + t_{ps} E_{{\rm in}}^{{\rm (p)}}) \hat{e}_p^+ 
+ 
(t_{sp} E_{{\rm in}}^{{\rm (s)}} + t_{pp} E_{{\rm in}}^{{\rm (p)}}) \hat{e}_s^+
\big) 
e^{i\kv^+\cdot\rv}
\label{H>R}
\ea
Applying Eqs.~(\ref{E<R}), (\ref{E>R}), (\ref{H<R}) and (\ref{H>R}) to Eqs.~(\ref{BCs}), we obtain simultaneous equations for the Fresnel coefficients:
\begin{subequations}
\ba
&&\frac{c k_z}{\omega} E_{{\rm in}}^{{\rm(p)}} 
- \frac{c k_z}{\omega} 
\big( r_{sp} E_{{\rm in}}^{{\rm(s)}} + r_{pp} E_{{\rm in}}^{{\rm(p)}} \big) 
= 
\frac{c k_z}{\omega}
\big( t_{sp} E_{{\rm in}}^{{\rm(s)}} + t_{pp} E_{{\rm in}}^{{\rm(p)}} \big), 
\\
&&(1+r_{ss}) E_{{\rm in}}^{{\rm(s)}} + r_{ps} E_{{\rm in}}^{{\rm(p)}} 
= t_{ss} E_{{\rm in}}^{{\rm(s)}} + t_{ps} E_{{\rm in}}^{{\rm(p)}}, 
\\
&&- \frac{c k_z}{\omega} 
\big( t_{ss} E_{{\rm in}}^{{\rm(s)}} + t_{ps} E_{{\rm in}}^{{\rm(p)}} \big) 
- \Big(
- \frac{c k_z}{\omega} E_{{\rm in}}^{{\rm(s)}} 
+ \frac{c k_z}{\omega} 
\big( 
r_{ss} E_{{\rm in}}^{{\rm(s)}} 
+ r_{ps} E_{{\rm in}}^{{\rm(p)}} 
\big)
\Big)
\nonumber\\
&=&
\frac{4\pi}{c} 
\Big(
\sigma_{xx} 
\big( 
(1+r_{ss}) E_{{\rm in}}^{{\rm(s)}} + r_{ps} E_{{\rm in}}^{{\rm(p)}} 
\big) 
- \sigma_{xy} 
\big( 
\frac{k_z}{k} E_{{\rm in}}^{{\rm(p)}} 
- \frac{k_z}{k} 
\big( r_{sp} E_{{\rm in}}^{{\rm(s)}} + r_{pp} E_{{\rm in}}^{{\rm(p)}} \big) 
\big)
\Big),
\\
&&E_{{\rm in}}^{{\rm(p)}} 
+ \big( r_{sp} E_{{\rm in}}^{{\rm(s)}} + r_{pp} E_{{\rm in}}^{{\rm(p)}} \big) 
-  \big( t_{sp} E_{{\rm in}}^{{\rm(s)}} + t_{pp} E_{{\rm in}}^{{\rm(p)}} \big) 
\nonumber\\
&=& 
\frac{4\pi}{c} 
\Big( 
- \sigma_{xx}
\frac{k_z}{k} 
\big( r_{sp} E_{{\rm in}}^{{\rm(s)}} + (r_{pp} - 1) E_{{\rm in}}^{{\rm(p)}} \big) 
+
\sigma_{xy}
\big( 
(1+r_{ss}) E_{{\rm in}}^{{\rm(s)}} + r_{ps} E_{{\rm in}}^{{\rm(p)}} 
\big)
\Big)
\ea
\end{subequations}
We obtain $r_{pp}, r_{ps}, t_{pp}$ and $t_{ps}$ by solving the equations for the case $E_{{\rm in}}^{{\rm(p)}} \neq 0$ and $E_{{\rm in}}^{{\rm(s)}} = 0$. Similarly, we obtain $r_{ss}, r_{sp}, t_{ss}$ and $t_{sp}$ by solving the equations for the case $E_{{\rm in}}^{{\rm(p)}} = 0$ and $E_{{\rm in}}^{{\rm(s)}} \neq 0$. 
This yields 
\ba
r_{ss} &\!\!=\!\!& 
-\frac{1}{\Delta}
\big( 
\tsigma_{xx}^2 + \tsigma_{xy}^2
+ \frac{\omega}{c k_z} \tsigma_{xx} 
\big),
\,\,
r_{ps} = r_{sp} 
= \frac{1}{\Delta} \tsigma_{xy}, 
\nonumber\\
r_{pp} &\!\!=\!\!& 
\frac{1}{\Delta} 
\big( 
\tsigma_{xx}^2 + \tsigma_{xy}^2 
+ \frac{c k_z}{\omega} \tsigma_{xx}
\big), 
\nonumber\\
t_{ss} &\!\!=\!\!& 
\frac{1}{\Delta}
\big( 
1 + \frac{ck_z}{\omega} \tsigma_{xx}
\big),
\,\,
t_{ps} = - t_{sp} 
= \frac{1}{\Delta} \tsigma_{xy}, 
\nonumber\\
t_{pp} &\!\!=\!\!& 
\frac{1}{\Delta}
\big( 
1 + \frac{\omega}{ck_z} \tsigma_{xx}
\big),
\ea
where $\tsigma_{\mu\nu}(\omega) \equiv (2\pi/c) \sigma_{\mu\nu}(\omega)$, $k_z \equiv ((\omega/c)^2 - k_\perp^2)^{1/2}$, and $\Delta \equiv \tsigma_{xx}^2 + \tsigma_{xy}^2 + \big( \frac{\omega}{c k_z} + \frac{c k_z}{\omega} \big) \tsigma_{xx} + 1$. 

Next, we consider the boundary value problem for Fig.~\ref{fig:BCs2}(a), which is a mirror image of the one we just considered for Fig.~\ref{fig:BCs2}(b). 
Now, in the right subspace, we have the incident electric field 
\be
\Ev_{{\rm in}}(\rv) 
= 
\big( E_{{\rm in}}^{{\rm (s)}} \hat{e}_s^- + E_{{\rm in}}^{{\rm (p)}} \hat{e}_p^- \big) e^{i\kv^-\cdot\rv}
\ee 
and its reflected field 
\ba
\Ev_{{\rm R}} (\rv)
&=& \big( E_{{\rm R}}^{{\rm (s)}} \hat{e}_s^+ + E_{{\rm R}}^{{\rm (p)}} \hat{e}_p^+ \big) e^{i\kv^+\cdot\rv}
\nonumber\\
&=& 
\big(
(r_{ss}' E_{{\rm in}}^{{\rm (s)}} + r_{ps}' E_{{\rm in}}^{{\rm (p)}}) \hat{e}_s^+ 
+ 
(r_{sp}' E_{{\rm in}}^{{\rm (s)}} + r_{pp}' E_{{\rm in}}^{{\rm (p)}}) \hat{e}_p^+
\big) 
e^{i\kv^+\cdot\rv}
\ea
The total electric field is 
\be
\Ev^>(\rv) 
= 
\big( E_{{\rm in}}^{{\rm (s)}} \hat{e}_s^+ 
+ E_{{\rm in}}^{{\rm (p)}} \hat{e}_p^+ \big) 
e^{i\kv^+\cdot\rv}
+
\big(
(r_{ss}' E_{{\rm in}}^{{\rm (s)}} + r_{ps}' E_{{\rm in}}^{{\rm (p)}}) \hat{e}_s^- 
+ 
(r_{sp}' E_{{\rm in}}^{{\rm (s)}} + r_{pp}' E_{{\rm in}}^{{\rm (p)}}) \hat{e}_p^-
\big) 
e^{i\kv^-\cdot\rv}
\label{E>Rm}
\ee
In the left subspace, we have the transmitted field
\be
\Ev^<(\rv) = 
\big(
(t_{ss}' E_{{\rm in}}^{{\rm (s)}} + t_{ps}' E_{{\rm in}}^{{\rm (p)}}) \hat{e}_s^- 
+ 
(t_{sp}' E_{{\rm in}}^{{\rm (s)}} + t_{pp}' E_{{\rm in}}^{{\rm (p)}}) \hat{e}_p^- 
\big) 
e^{i\kv^-\cdot\rv}
\label{E<Rm}
\ee
In the expressions above, we have introduced primed notation for the Fresnel coefficients, to remind ourselves that these are for the left interface, which is a mirror image of the right interface. 
The magnetic field in the right subspace is now given by 
\ba
\Hv^> &=& -i(c/\omega) {{\bm \nabla}} \times \Ev^>
\nonumber\\
&=&
\frac{c}{\omega}
\big( E_{{\rm in}}^{{\rm (s)}} \kv^- \times \hat{e}_s^- 
+ E_{{\rm in}}^{{\rm (p)}} \kv^- \times \hat{e}_p^- \big) 
e^{i\kv^-\cdot\rv}
\nonumber\\
&&+
\frac{c}{\omega}
\big(
(r_{ss}' E_{{\rm in}}^{{\rm (s)}} + r_{ps}' E_{{\rm in}}^{{\rm (p)}}) \kv^+ \times \hat{e}_s^+ 
+ 
(r_{sp}' E_{{\rm in}}^{{\rm (s)}} + r_{pp}' E_{{\rm in}}^{{\rm (p)}}) \kv^+ \times \hat{e}_p^+
\big) 
e^{i\kv^+\cdot\rv}
\nonumber\\
&=&
\frac{ck}{\omega} 
\big(
- E_{{\rm in}}^{{\rm (s)}} \hat{e}_p^- 
+ E_{{\rm in}}^{{\rm (p)}} \hat{e}_s^-  
\big) 
e^{i\kv^-\cdot\rv}
+
\frac{ck}{\omega}
\big(
- (r_{ss}' E_{{\rm in}}^{{\rm (s)}} + r_{ps}' E_{{\rm in}}^{{\rm (p)}}) \hat{e}_p^+ 
+ 
(r_{sp}' E_{{\rm in}}^{{\rm (s)}} + r_{pp}' E_{{\rm in}}^{{\rm (p)}}) \hat{e}_s^+
\big) 
e^{i\kv^+\cdot\rv}
\label{H>Rm}
\ea
For the magnetic field in the left subspace, we find
\ba
\Hv^< &=& -i(c/\omega) {{\bm \nabla}} \times \Ev^<
\nonumber\\
&=& 
\frac{c}{\omega}
\big(
(t_{ss}' E_{{\rm in}}^{{\rm (s)}} + t_{ps}' E_{{\rm in}}^{{\rm (p)}}) (\kv^- \times \hat{e}_s^-) 
+ 
(t_{sp}' E_{{\rm in}}^{{\rm (s)}} + t_{pp}' E_{{\rm in}}^{{\rm (p)}}) (\kv^- \times \hat{e}_p^-)
\big) 
e^{i\kv^-\cdot\rv}
\nonumber\\
&=&
\frac{ck}{\omega}
\big(
- (t_{ss}' E_{{\rm in}}^{{\rm (s)}} + t_{ps}' E_{{\rm in}}^{{\rm (p)}}) \hat{e}_p^- 
+ 
(t_{sp}' E_{{\rm in}}^{{\rm (s)}} + t_{pp}' E_{{\rm in}}^{{\rm (p)}}) \hat{e}_s^- 
\big) 
e^{i\kv^-\cdot\rv}
\label{H<Rm}
\ea
Substituting Eqs.~(\ref{E>Rm}), (\ref{E<Rm}), (\ref{H>Rm}) and (\ref{H<Rm}) into Eqs.~(\ref{BCs}) leads to 
\begin{subequations}
\ba
&&- \frac{c k_z}{\omega} E_{{\rm in}}^{{\rm(p)}} 
+ \frac{c k_z}{\omega} 
\big( r_{sp}' E_{{\rm in}}^{{\rm(s)}} + r_{pp}' E_{{\rm in}}^{{\rm(p)}} \big) 
= 
- \frac{c k_z}{\omega}
\big( t_{sp}' E_{{\rm in}}^{{\rm(s)}} + t_{pp}' E_{{\rm in}}^{{\rm(p)}} \big), 
\\
&&(1+r_{ss}') E_{{\rm in}}^{{\rm(s)}} + r_{ps}' E_{{\rm in}}^{{\rm(p)}} 
= t_{ss}' E_{{\rm in}}^{{\rm(s)}} + t_{ps}' E_{{\rm in}}^{{\rm(p)}}, 
\\
&&- \frac{c k_z}{\omega} 
\big( t_{ss}' E_{{\rm in}}^{{\rm(s)}} + t_{ps}' E_{{\rm in}}^{{\rm(p)}} \big) 
+ \Big(
 \frac{c k_z}{\omega} E_{{\rm in}}^{{\rm(s)}} 
- \frac{c k_z}{\omega} 
\big( 
r_{ss}' E_{{\rm in}}^{{\rm(s)}} 
+ r_{ps}' E_{{\rm in}}^{{\rm(p)}} 
\big)
\Big)
\nonumber\\
&=&
\frac{4\pi}{c} 
\Big(
\sigma_{xx} 
\big( 
(1+r_{ss}') E_{{\rm in}}^{{\rm(s)}} + r_{ps}' E_{{\rm in}}^{{\rm(p)}} 
\big) 
- \sigma_{xy} 
\big( 
-  \frac{k_z}{k} E_{{\rm in}}^{{\rm(p)}} 
+ \frac{k_z}{k} 
\big( r_{sp}' E_{{\rm in}}^{{\rm(s)}} + r_{pp}' E_{{\rm in}}^{{\rm(p)}} \big) 
\big)
\Big),
\\
&&- E_{{\rm in}}^{{\rm(p)}} 
- \big( r_{sp}' E_{{\rm in}}^{{\rm(s)}} + r_{pp}' E_{{\rm in}}^{{\rm(p)}} \big) 
+ \big( t_{sp}' E_{{\rm in}}^{{\rm(s)}} + t_{pp}' E_{{\rm in}}^{{\rm(p)}} \big) 
\nonumber\\
&=& 
\frac{4\pi}{c} 
\Big( 
\sigma_{xx}
\frac{k_z}{k} 
\big( r_{sp}' E_{{\rm in}}^{{\rm(s)}} + (r_{pp}' - 1) E_{{\rm in}}^{{\rm(p)}} \big) 
+
\sigma_{xy}
\big( 
(1+r_{ss}') E_{{\rm in}}^{{\rm(s)}} + r_{ps}' E_{{\rm in}}^{{\rm(p)}} 
\big)
\Big)
\ea
\label{B17}
\end{subequations}
Again, we can obtain the Fresnel coefficients by solving separately for the cases when the incident ray is s-polarized and when it is p-polarized. This leads to $r_{ss}' = r_{ss}$, $r_{pp}' = r_{pp}$, $t_{ss}' = t_{ss}$, $t_{pp}' = t_{pp}$, and $r_{ps}' = -r_{ps}$, $r_{sp}' = -r_{sp}$, $t_{ps}' = -t_{ps}$ and $t_{sp}' = -t_{sp}$. The diagonal reflection coefficients of $\Rv'$ are thus the same as those of $\Rv$, but the off-diagonal reflection coefficients of the two reflectivity matrices differ by a sign. In the regime of large separations, the Casimir-Lifshitz energy of two Chern insulators (labeled 1 and 2) is dominated by the contributions of the off-diagonal entries of the reflectivity matrices $\Rv_1'$ and $\Rv_2$. The different signs of the off-diagonal entries of $\Rv_1'$ and $\Rv_2$ leads to the positive sign factor in front of the energy in Eq.~(\ref{ECI}).


\begin{thebibliography}{999}
\bibitem{footnote:largeaxion}
\bing{For larger values of $\bar{\alpha}$ the force reverts to attraction. This is analogous to what happens for Chern insulators that have large values of the Chern number, which we discuss in Sec.~\ref{sec:chern}.}
\bibitem{footnote:bigC}
Note that the result mentioned here, i.e., the Chern insulators attract each other as perfect conductors in the limit that $|C|\alpha \to \infty$, is correct. It is different from that in Ref.~\cite{rodriguez-lopez2014}, where they found that the energy is equal to $0.92$ times that of a pair of perfect conductors. The author thanks an anonymous referee for pointing this out.
\end{thebibliography}
\end{document}